\documentclass[prb,amsmath,amssymb,twocolumn,lengthcheck,showpacs,floatfix,groupedaddress,superscriptaddress,longbibliography]{revtex4-1}
\usepackage{graphicx}
\usepackage[english]{babel}
\usepackage{times}
\usepackage{dcolumn}
\usepackage[usenames,dvipsnames]{color}
\usepackage{units}
\usepackage{bm}

\begin{document}

\title{Repulsive vs. attractive Hubbard model: transport and dynamical
properties} 

\author{Rok \v{Z}itko}

\affiliation{Jo\v{z}ef Stefan Institute, Jamova 39, SI-1000 Ljubljana, Slovenia}
\affiliation{Faculty  of Mathematics and Physics, University of Ljubljana, 
Jadranska 19, SI-1000 Ljubljana, Slovenia}

\author{\v{Z}iga Osolin}

\affiliation{Jo\v{z}ef Stefan Institute, Jamova 39, SI-1000 Ljubljana, Slovenia}

\author{Peter Jegli\v{c}}

\affiliation{Jo\v{z}ef Stefan Institute, Jamova 39, SI-1000 Ljubljana,
Slovenia}

\date{\today}

\begin{abstract}
We contrast the transport properties (dc resistivity, Seebeck
coefficient), optical conductivity, spectral functions, dynamical
magnetic susceptibility, and the NMR $1/T_1$ spin-lattice relaxation
rate of the repulsive and attractive infinite-dimensional Hubbard
models in the paramagnetic phase for a generic band filling. The
calculations are performed in a wide temperature interval using the
dynamical mean-field theory with the numerical renormalization group
as the impurity solver. The attractive case exhibits significantly
more complex temperature dependences which can be explained by the
behavior of the half-filled Hubbard model in external magnetic field
with constant magnetization, to which the attractive Hubbard model
maps through the partial particle-hole transformation. The resistivity
is non-monotonous for strongly attractive case: it peaks significantly
above the MIR value at a temperature $T_\mathrm{max}$ where the
quasiparticle band disappears. For both signs of $U$ we find
particle-hole asymmetry in the self-energy at low energies, but with
the opposite kind of excitations having longer lifetime. This leads to
a strong suppression of the slope of the Seebeck coefficient in the
attractive case, rather than an enhancement as in the repulsive case.
The spin-lattice relaxation rate in the strongly attractive case has a
non-monotonic temperature dependence, thereby revealing the pairing
fluctuations.
\end{abstract}


\maketitle

\newcommand{\vc}[1]{{\mathbf{#1}}}
\renewcommand{\Im}{\mathrm{Im}}
\renewcommand{\Re}{\mathrm{Re}}
\newcommand{\expv}[1]{\langle #1 \rangle}
\newcommand{\beq}[1]{\begin{equation} #1 \end{equation}}
\newcommand{\corr}[1]{\langle\langle #1 \rangle\rangle}

\newcommand{\braket}[2]{\langle#1|#2\rangle}
\newcommand{\ket}[1]{| #1 \rangle}
\newcommand{\Tr}{\mathrm{Tr}}
\newcommand{\degg}{^\circ}
\renewcommand{\Im}{\mathrm{Im}}
\renewcommand{\Re}{\mathrm{Re}}
\newcommand{\GG}{{\mathcal{G}}}
\newcommand{\atanh}{\mathrm{atanh}}
\newcommand{\sgn}{\mathrm{sgn}}

\section{Introduction}

Electrons in materials are charged particles that repel each other
through Coulomb interaction, but effective electron-electron
attraction can be generated by coupling to lattice vibrations
\cite{cooper1956}. The Hubbard model \cite{hubbard1963, kanamori1963,
gutzwiller1963} describes a lattice with a short-ranged (on-site)
electron-electron interaction $U$ which can be either positive
(repulsion) or negative (attraction). The repulsive Hubbard model
is a minimal model for the cuprate family of superconducting materials
\cite{scalapino1995,lee2006} and describes the competition between the
delocalizing effects of electron hopping and localizing effects of
charge repulsion. 
The attractive Hubbard model is used as an effective description for
certain systems with very strong electron-phonon coupling and for cold
atoms in optical lattices
\cite{moreo2007,Hackermuller:2010cm,Schneider:2012ke}. It has been
used to study, for example, strong-coupling superconductors and the
continuous cross-over between the BEC and BCS superconducting regimes
\cite{micnas1990,keller2001,capone2002,toschi2005pairing,chien2008,bauer2009,kuleeva2014}.

There are very few works that directly address the differences between
the repulsive and the attractive regime of the Hubbard model. While at
the particle-hole symmetric point (i.e., at half filling, for one
electron per lattice site), the two cases are trivially related by a
partial particle-hole transformation that leads to $U \to -U$ and
simply exchanges the spin and charge sectors, this is no longer the
case at finite doping, since the doping corresponds to the
magnetization under this mapping. Comparative studies of repulsive and
attractive Hubbard models are very valuable for understanding more
complex models such as the Hubbard-Holstein model
\cite{Koller:2004ic,paci2006,Bauer:2010hx,Bauer:2010by}, where for
increasing electron-phonon (e-ph) coupling the effective
electron-electron (e-e) interaction becomes attractive on low-energy
scales, while remaining repulsive at higher energies. They are also of
interest in the context of fermionic cold atoms trapped in optical
lattices \cite{Georges:2007vs}, where the strength and even the sign
of the interaction can be tuned by means of Feshbach resonances. In
this work, we study the paramagnetic phase of the Hubbard model at
moderate hole doping, $\expv{n}=0.8$, for both signs of $U$ using the
dynamical mean-field theory (DMFT) \cite{metzner1989,georges1996}. 
Magnetic order, charge-density-wave, and superconducting DMFT
solutions are also possible
\cite{jarrell1992dmft,georges1993,rozenberg1994,keller2001,capone2002,toschi2005pairing,bauer2009,Koga:2011hj},
but not considered in our calculations. In other words, we only
consider the paramagnetic (nonmagnetic, normal-state) phase that is
uniform in space. Even if the true ground state is actually ordered,
our results are still valid above the ordering temperature
\cite{keller2001,capone2002,toschi2005pairing,georges2011}.
Furthermore, since the ordering temperatures can be significantly
reduced by frustration (such as that due to the next-nearest-neighbour
hopping or external magnetic field), the range of qualitative validity
of our results can extend to very low temperatures in such cases
\cite{keller2001,toschi2005pairing}.

We focus on the experimentally most relevant properties: transport
(resistivity and Seebeck coefficient), optical conductivity, and NMR
$1/T_1$ spin-lattice relaxation rate as a function of temperature, but
we also provide results for thermodynamics, spectral functions, and
dynamical spin susceptibility. The main new results of this work
concern the attractive Hubbard model: (a) identification of the
characteristic energy scales, 
(b) opposite signs of the particle-hole asymmetry of velocities
and scattering rate, leading to a near-cancellation of the
contributions to the low-temperature Seebeck coefficient, and (c) the
non-monotonic temperature dependence of the spin-lattice relaxation
rate.

This work is structured as follows. In Sec.~\ref{sec2} we introduce
the model and discuss the partial particle-hole transformation. In
Sec.~\ref{sec3} we describe the thermodynamic properties as a function
of Hubbard coupling $U$ and temperature $T$. In Sec.~\ref{sec4} we
discuss the local and momentum-resolved spectral functions, the
$U$-dependence of the quasiparticle renormalization factor $Z$, and
the asymmetric structure of the self-energy $\Sigma$ and its
temperature variation. In Sec.~\ref{sec5} we describe the transport
properties and provide some details about the non-monotonous
temperature dependences in the attractive Hubbard model. In
Sec.~\ref{sec6} we compare the spin-lattice relaxation rates and
discuss the temperature dependence of the dynamical susceptibilities.
Section~\ref{sec7} is devoted to the DMFT mapping in the attractive
$U$ case, where the effective model is the particle-hole symmetric
Anderson impurity model at constant magnetization and we discuss to
what degree the properties of the impurity model reflect in the fully
self-consistent DMFT calculations. The final section~\ref{sec8}
concerns the experimental relevance of our calculations and presents
some additional results for the optical conductivity that could aid in
the interpretation of the measurements on zeolite materials.

\section{Model and method}
\label{sec2}

We study the Hubbard model 
\beq{
H=\sum_{\vc{k}\sigma} \epsilon_\vc{k} c^\dag_{\vc{k}\sigma}
c_{\vc{k}\sigma} + U \sum_i n_{i\uparrow} n_{i\downarrow}.
}
$\epsilon_\vc{k}$ is the dispersion relation of electrons with
wave-vector $\vc{k}$ and spin $\sigma$, $U$ is the Hubbard coupling.
Index $i$ ranges over all lattice sites, while $n_{i\sigma}=
c^\dag_{i\sigma} c_{i\sigma}$. 

We seek a non-ordered solution of this model using the DMFT
\cite{metzner1989, mullerhartmann1989, georges1996}. In this approach,
the bulk problem defined on the lattice maps onto a quantum impurity
model (here the single impurity Anderson model) subject to a
self-consistency condition for the hybridization function
\cite{georges1992, rozenberg1992, jarrell1992dmft, zhang1993}. This
technique takes into account all local quantum fluctuations exactly,
while the inter-site correlations are treated at the static mean-field
level. This is a good approximation for problems where the most
important effects are local in nature (Mott metal-insulator
transition, etc.). It is an exact method in the limit of infinite
dimensions or infinite lattice connectivity, and appears to be
reasonably reliable as an approximative technique for 3D lattices
\cite{georges1996,kotliar2006}, while for 2D and 1D systems it is less
applicable due to stronger non-local fluctuations.

We work with the Bethe lattice that has non-interacting density of
states (DOS)
\begin{equation}
\rho_0(\epsilon) = \frac{2}{\pi D} \sqrt{1-(\epsilon/D)^2},
\end{equation}
which mimics some of the features of the 3D-cubic lattice DOS, in
particular the square root band-edge singularities. $D$ is the
half-bandwidth that we use to express the parameters and the results
as dimensionless quantities.

As the impurity solver, we use the numerical renormalization group
(NRG) \cite{wilson1975, krishna1980a, sakai1989, costi1994, bulla1998,
bulla1999, pruschke2000, bulla2008, resolution} with discretization
parameter $\Lambda=2$, twist averaging over $N_z=16$ values
\cite{frota1986,campo2005}, and keeping up to 12000 multiplets (or up
to a truncation cutoff at energy $10\omega_N$, where $\omega_N$ is the
characteristic energy at the $N$-th NRG step). The twist averaging in
the NRG means that $N_z$ separate NRG calculations are run for
different choices of interleaved discretization grids (so-called $z$
parameters) and the results are then averaged; this technique leads to
a significant cancelation of the discretization artifacts of the
method. Spectral broadening has been performed with parameter
$\alpha=0.3$. We use Broyden's method to speed-up the convergence of
the DMFT iteration and to control the chemical potential in the
constant-occupancy calculations \cite{broyden}. The convergence
criteria are very stringent (integrated absolute value of the
difference of spectral functions less than $10^{-8}$) in an attempt to
obtain reliable results for transport properties at low temperatures.
In spite of these efforts, the residual oscillatory features in the
self-energy remain problematic at low temperatures; for computing
transport properties it is necessary to perform fitting of the
self-energy with low-order polynomials around $\omega=0$. In
particular, the results for the Seebeck coefficient turn out to be
exceedingly difficult to compute reliably at very low temperatures.

On bipartite lattices the repulsive and the attractive Hubbard models
are related through the partial particle-hole (Lieb-Mattis)
transformation \cite{lieb1989,moreo2007,capone2002,toschi2005pairing}
defined as
\beq{
c^\dag_{i\uparrow} \to d^\dag_{i\uparrow},
\quad
c^\dag_{i\downarrow} \to (-1)^i d_{i\downarrow}.
}
For down spins, this can be interpreted as a mapping of the particle
creation operators onto the annihilation operators for the holes. The
$(-1)^i$ factor indicates different prefactors for the two sublattices
of a bipartite lattice. The transformation leaves the kinetic energy
unchanged, but changes the sign of the quartic electron-electron
coupling term, i.e., flips the sign of $U$. Furthermore, it can be
seen that the particle number (density) operator for $c$ particles
maps onto the spin-$z$ (magnetization) operator for $d$ particles.
While the spin-up Green's function is invariant, the spin-down Green's
function is transformed. Since $\corr{A ; B}_{z} = -\corr{B ;
A}_{-z}$, the transformation is
\beq{
\label{Amap}
A_{i\downarrow}(\omega) \to A_{i\downarrow}(-\omega).
}
This implies that the field-induced Zeeman splitting of the
quasiparticle band in the $U>0$ case corresponds to a uniform shift of
the quasiparticle band through changes of the chemical potential in
the $U<0$ case. This has imporant consequences for the transport
properties, especially for the Seebeck coefficient which is sensitive
to the particle-hole asymmetry.

Unless noted otherwise, the band filling is $\expv{n}=0.8$, i.e., the
hole doping level is $\delta=1-\expv{n}=0.2$, which is sufficiently
away from any special points to be considered as a generic band
filling. For attractive $U$, similar DMFT studies have been performed
using different impurity solvers (Hirsch-Fye QMC, exact
diagonalization), focusing on the pairing transition in the
paramagnetic case \cite{keller2001,capone2002} and on the
superconducting solution \cite{toschi2005pairing}. The advantage of
the NRG compared to those works is in the higher spectral resolution
and large temperature range of applicability, from $T=0$ to
temperatures comparable to the bandwidth. Some results for the
attractive $U$ computed using the DMFT(NRG) approach have recently
been reported \cite{kuleeva2014}.

The attractive Hubbard model on the infinite-connectivity Bethe
lattice (and more generally on bipartite lattices in dimension higher
than two) has a superconducting solution for all $U$ and all densities
$n$ \cite{micnas1990}. If the superconductivity is suppressed, the
normal-state is a Fermi liquid (metallic) for $U > U_0$ and a
bound-pair (insulating) state for $U < U_0$, separated by a pairing
quantum phase transition at $U_0$ which is equivalent to the Mott
metal-insulator-transition in the presence of the magnetic field for
the $U>0$ model
\cite{keller2001,capone2002,toschi2005pairing,laloux1994,georges1996,bauer2007fm,bauer2009fm,vanhove}.
For finite doping, it has been shown that the transition is first
order \cite{capone2002}.

\section{Thermodynamic properties}
\label{sec3}

\begin{figure}
\centering
\includegraphics[clip,width=0.48\textwidth]{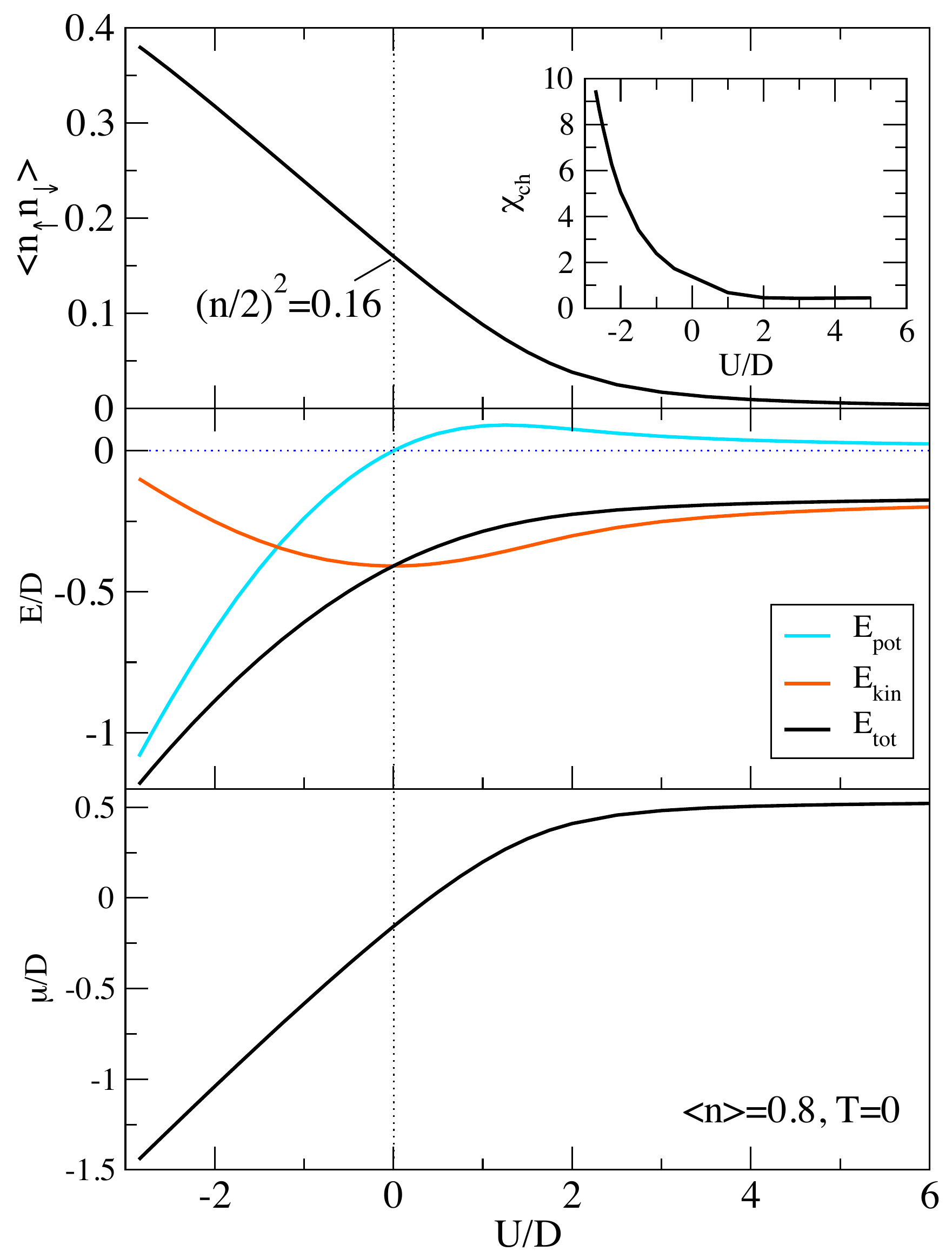}
\caption{(Color online) Zero-temperature thermodynamic properties of
the Hubbard model as a function of the electron-electron interaction
parameter $U$. The density is fixed at $n=0.8$. (a) Double occupancy
(density of doubly occupied sites) $P_2=\expv{n_\uparrow
n_\downarrow}$. The inset shows the uniform charge susceptibility,
$\chi_\mathrm{c}=\partial \expv{n}/\partial \mu$. The temperature
depdendce of $P_2$ is shown in Fig.~\ref{NMRT}. (b) Potential,
kinetic, and total energy per particle. (c) Chemical potential.}
\label{fig1}
\end{figure}

We first consider the static (thermodynamic) properties. In
Fig.~\ref{fig1}(a) we show the double occupancy $P_2=\expv{n_\uparrow
n_\downarrow}$, which is a measure of local pair formation
\cite{keller2001}. The non-interacting result at $U=0$,
$(n/2)^2=0.16$, is rapidly reduced for repulsive $U$ with maximum
curvature in the range where the upper Hubbard band emerges ($U
\approx 2D$, see Fig.~\ref{fig2}) and tends to zero as $1/U$ in the
large-$U$ limit. For attractive $U$, the double occupancy at zero
temperature increases with increasing $|U|$ up to values close to
$n/2=0.4$, at which point the constant-occupancy DMFT calculations no
longer converge due to a very high charge susceptibility close to the
pairing phase transition (see the inset in Fig.~\ref{fig1}) and the
coexistence of several solutions of the DMFT equations
\cite{capone2002}. Asymptotically, in the pairing phase, one would
expect that all particles are bound as local pairs for infinite
attraction, so that $P_2\to n/2=0.4$ when $U \to -\infty$. In the
parameter range where $P_2$ becomes large and the convergence slow, it
helps to perform the DMFT calculations at a fixed chemical potential
$\mu$ and determine the appropriate $\mu$ by bisection; this becomes
crucial in the parameter range where there is a phase separation. 
The instability also manifests itself as a large spread of the
expectation values of physical observables in the $z$-averaging method
in the NRG calculations. For example, at $U/D=-2.85$, the computed
$\expv{n}$ values range from 0.741 to 0.861 for different
discretization grids, thus the quantitative validity of the results
becomes questionable (for comparison, generally the differences
between $\expv{n}$ are of order $10^{-4}$). Such behavior is a well
known precursor of phase transitions in the NRG calculations. The NRG
calculations using the twist averaging must namely be performed with
caution close to quantum phase transitions, since for different values
of $z$ the system may be in different phases, thus the $z$ averaging
itself becomes meaningless. The severity of this problem depends on
the system and on the type of the transition. For the attractive
Hubbard model constrained to the normal phase, as studied here, it the
difficulties are particularly strong. Therefore, using the DMFT(NRG)
approach it is difficult to locate the transition point and to study
its nature. \footnote{One way to proceed is to perform the NRG
calculation for different values of $z$ {\it without} averaging the
results, thus obtaining different transition points for the different
values of $z$. The true transition occurs at the average of the
$z$-dependent values. In addition, in such cases it becomes important
to consider the dependence of the results on the discretization
parameter $\Lambda$ and the approach to the continuum limit $\Lambda
\to 1$.} 

In Fig.~\ref{fig1}(b) we follow the kinetic and potential
energies. The potential energy is given simply by $U \expv{n_\uparrow
n_\downarrow}$, thus it does not bring any new information.
$E_\mathrm{kin}$ is minimal in the non-interacting case. It increases
for both signs of $U$, because interactions of both signs lead to
increased particle localization which costs kinetic energy. 

\begin{figure}
\centering
\includegraphics[clip,width=0.48\textwidth]{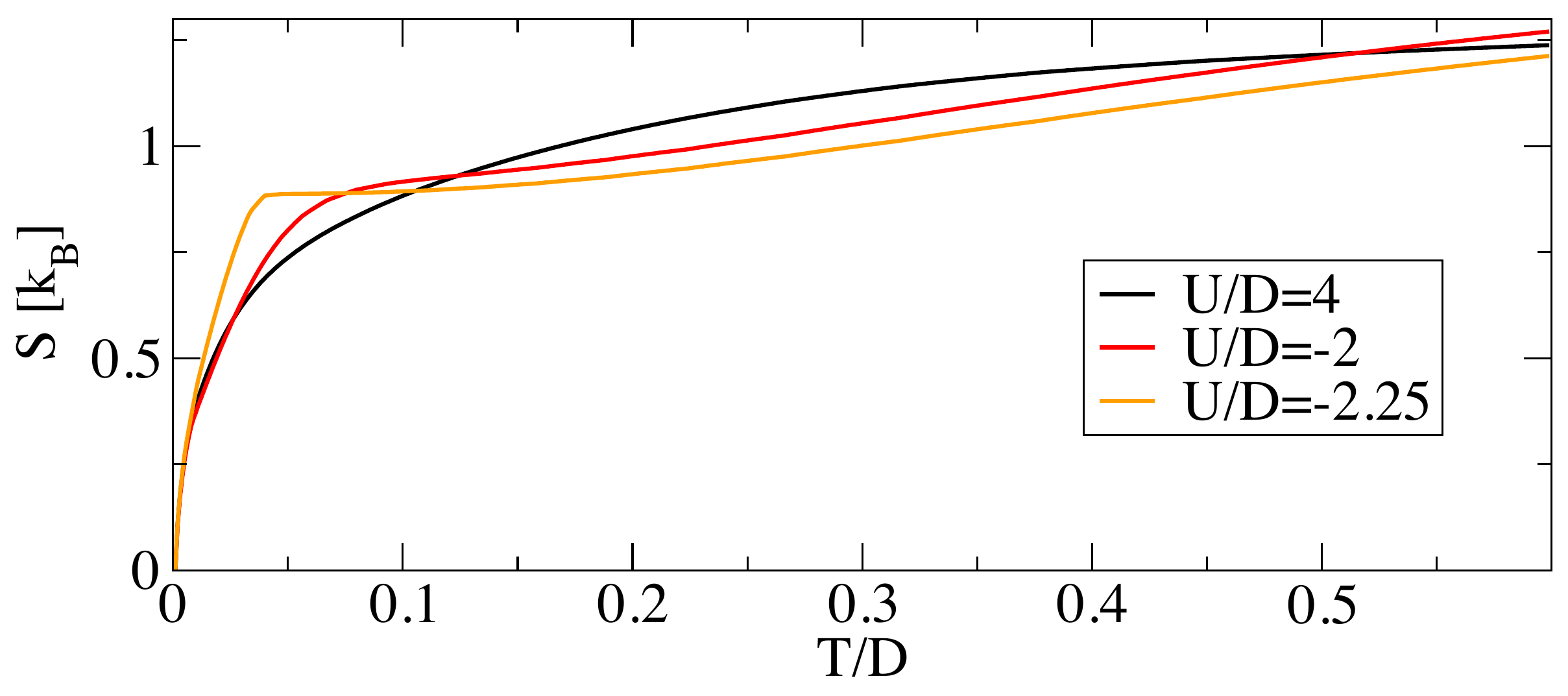}
\caption{(Color online) Temperature dependence of entropy per lattice
site for a set of $U$ values.
}
\label{fig1b}
\end{figure}

\begin{figure}
\centering
\includegraphics[clip,width=0.48\textwidth]{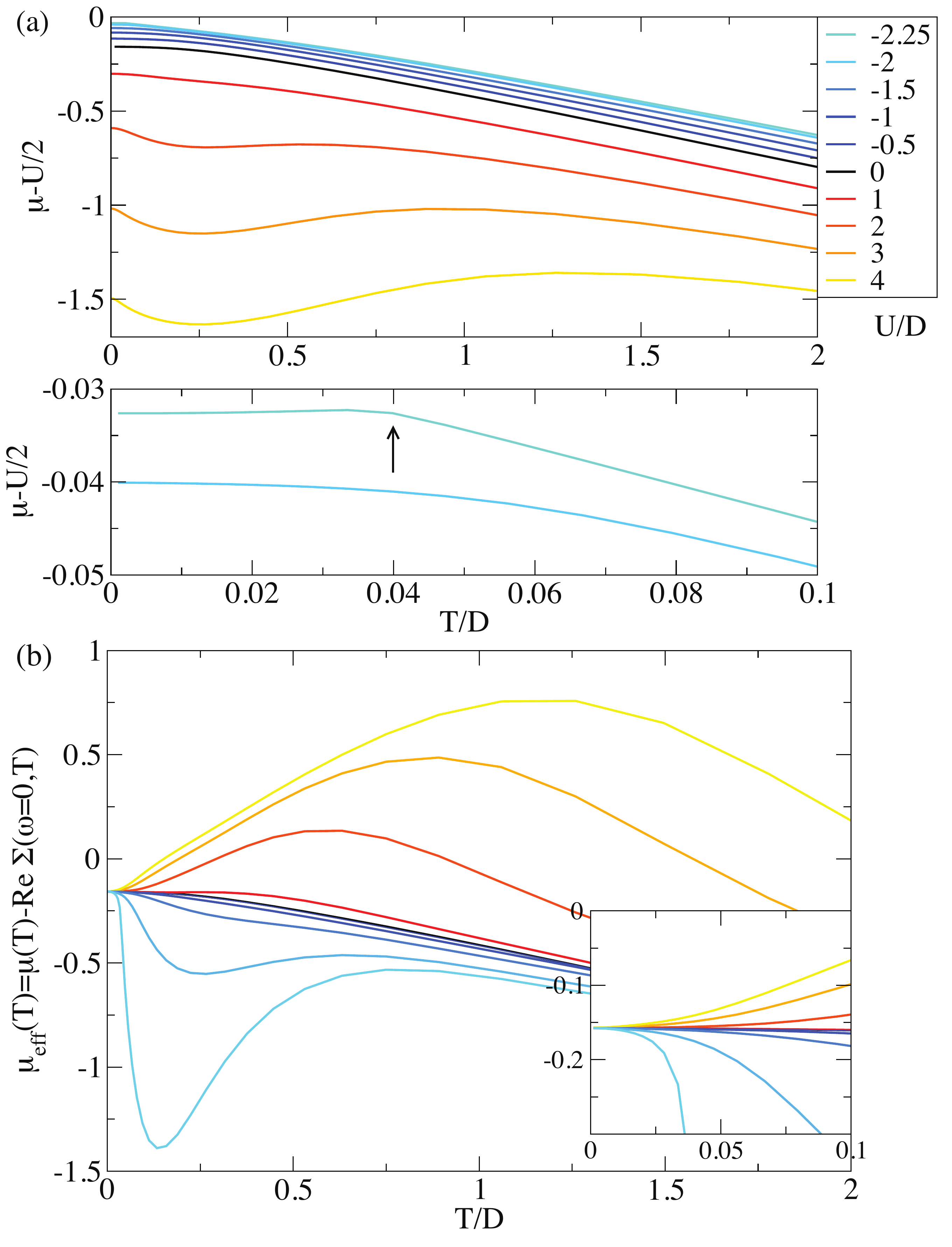}
\caption{(Color online) (a) Temperature dependence of the chemical
potential (shifted by $-U/2$). The zero-temperature values of the
chemical potential (without the shift) are also shown in
Fig.~\ref{fig1}c. (b) Temperature dependence of the renormalized
chemical potential
$\mu_\mathrm{eff}(T)=\mu(T)-\mathrm{Re}\Sigma(\omega=0,T)$.}
\label{fig33}
\end{figure}

We now consider the temperature dependence of the entropy. In
Fig.~\ref{fig1b} we show representative cases for strongly repulsive
and strongly attractive interaction. For both signs of $U$, the
entropy attains values of order $\ln 2 \approx 0.69$ already at
relatively low temperatures. This indicates the presence of
fluctuating local moments (for repulsive $U$, i.e., a bad metal regime
of doped Mott insulators) or paired states (attractive $U$, i.e., an
incoherent pairing state). The entropy curves for attractive $U$ have
a pronounced plateau at intermediate temperatures. For example, at
$U/D=-2.25$ the low-temperature nearly linear region is followed by a
plateau starting at $T=T_\mathrm{pl} \approx 0.04D$,
up to $T \approx 0.1D$ at which point
it starts to gradually rise again. The temperature scale
$T_\mathrm{pl}$ is also visible in the chemical potential $\mu(T)$:
for $T<T_\mathrm{pl}$ the chemical potential is nearly constant, then
it rapidly crosses-over into a new decreasing regime that smoothly
connects with the asymptotic linear behavior, see Fig.~\ref{fig33}. At
$T_\mathrm{pl}$ the quasiparticle band is already reduced, but not yet
fully eliminated.

We note that under the $U\to-U$ mapping, the chemical potential
corresponds to the magnetic field required to maintain the
magnetization constant. In the following, we show that the plateau
starting at $T_\mathrm{pl}$ can be related to features seen in the
double occupancy, dynamical susceptibility, and spin-lattice
relaxation curves, but not so well in the transport properties. It
indicates the regime where the electron pairing interaction tends to
eliminate the coherent Fermi liquid state.

In Fig.~\ref{fig1}(b) we show the temperature dependence of the
renormalized chemical potential defined as
\begin{equation}
\mu_\mathrm{eff}(T)=\mu(T)-\mathrm{Re}\Sigma(\omega=0,T).
\end{equation}
This quantity determines the location of the peak in the momentum
distribution curves $A(\epsilon,\omega=0)$. At $T=0$, its value is
fixed by the Luttinger theorem to the non-interacting Fermi level. For
strong interaction of either sign, the renormalized chemical potential
deviates strongly from the $U=0$ result already at very low
temperatures on the scale of $T_F$. For repulsive interaction, as the
temperature increases the Fermi volume first expands \cite{deng} (in
the sense that the peak in the momentum distribution shifts to higher
$\epsilon$ at higher temperatures), while for attractive interaction
it {\sl contracts}. This provides a simple picture: the repulsive
interaction tends to expand the Fermi sphere upon heating (electrons
reduce double occupancy of the occupied $\epsilon_k$ levels), while
the attractive interaction contracts it (electrons increase double
occupancy of the occupied $\epsilon_k$ levels); this is also confirmed
by the temperature dependence of pairing, show in Fig.~\ref{NMRT}(b).
For repulsive interaction, this trend continues to high temperatures
and reverses on a scale determined by $U$ where the system approaches
the atomic limit. For attractive interaction, the Fermi surface
contraction terminates on an intermediate temperature scale of order
$ZD$; this is followed up by a region of increasing $\mu_\mathrm{eff}$
until the final approach to the atomic limit where $\mu_\mathrm{eff}$
is decreasing.

\section{Single-particle dynamical properties}
\label{sec4}

\subsection{Zero-temperature spectral functions}

\begin{figure*}
\centering
\includegraphics[clip,width=0.9\textwidth]{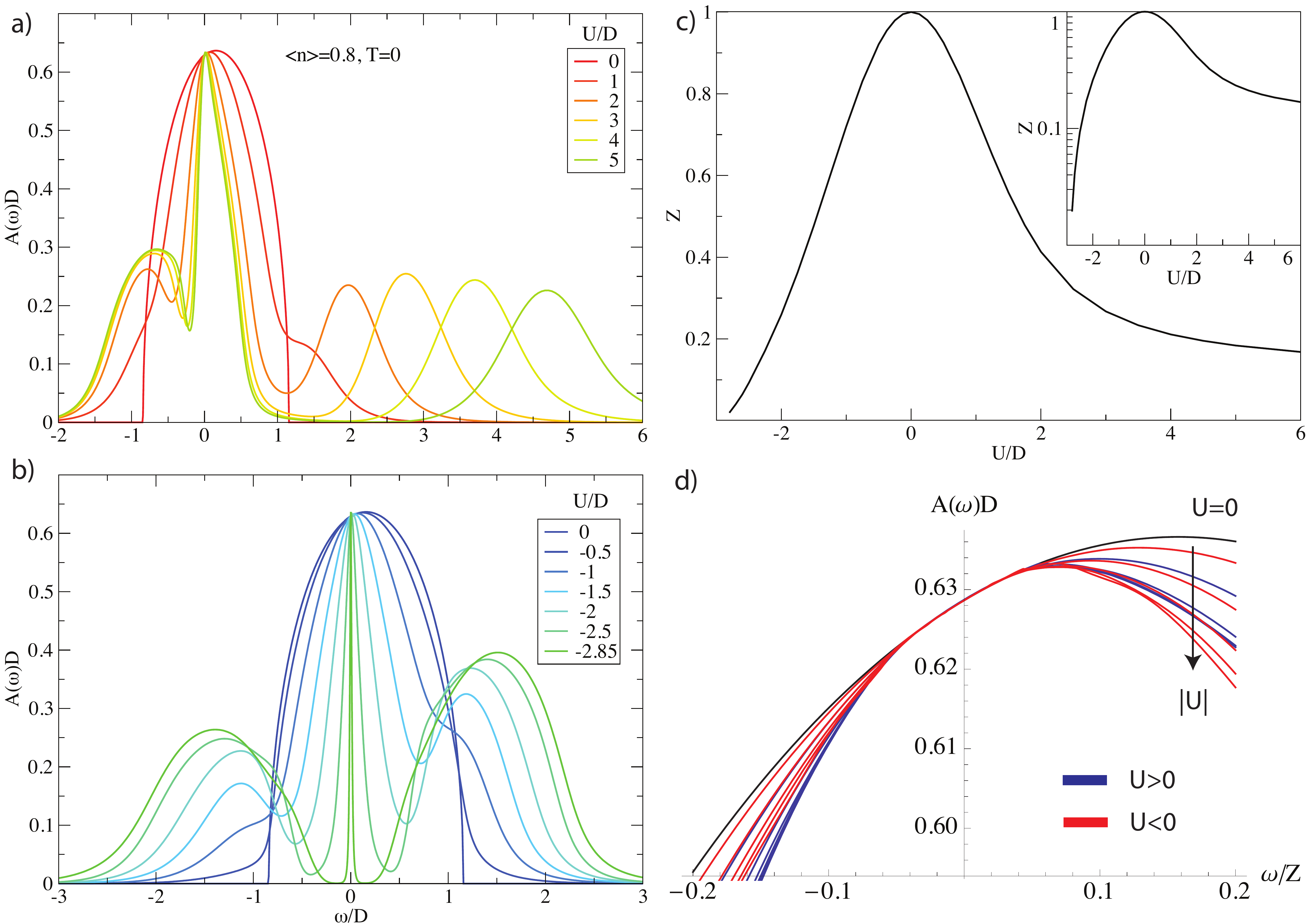}
\caption{(Color online) Local spectral function $A(\omega)$ at zero
temperature for (a) repulsive and (b) attractive case.  (c)
Quasiparticle renormalization factor $Z\equiv Z(T=0)$ as a function of
$U$. 
(d) Low-frequency
part of the spectral function rescaled as $A(\omega/Z)$. We plot the
results for $U=-2.5,-2,-1.5,-1,-0.5,0,1,2,3,4,5,6D$. }
\label{fig2}
\end{figure*}

In the DMFT, the lattice (momentum-resolved) Green's function is
approximated using a self-energy function that depends only on
the frequency but not on the momentum, so that
\beq{
G_\vc{k}(z) = \frac{1}{z+\mu-\epsilon_\vc{k}-\Sigma(z)},
}
where $z$ is complex frequency (one may take $z=\omega+i\delta$ to
obtain the retarded Green's function). The local Green's function is
obtained as the $\vc{k}$ average:
\beq{
\begin{split}
G_\mathrm{loc}(z) &= \frac{1}{N} \sum_\vc{k} G_\vc{k}(z)
= \int \frac{\rho_0(\epsilon)\,d\epsilon}{z+\mu-\epsilon-\Sigma(z)} \\
&= G_0[z+\mu-\Sigma(z)],
\end{split}
}
where $N$ is the number of lattice sites and $G_0(z)$ is the
non-interacting Green's function of the chosen lattice,
here
\begin{equation}
G_0(z) = \frac{2}{D} \left( z/D - \mathrm{sign}[\Im(z)] \sqrt{1-(z/D)^2} \right).
\end{equation}
Momentum-resolved and local spectral functions are then defined as
$A_\vc{k}(\omega)=(-1/\pi) \Im G_\vc{k}(\omega+i\delta)$ and
$A(\omega)=(-1/\pi)\Im G_\mathrm{loc}(\omega+i\delta)$.

In Fig.~\ref{fig2}(a) and (b) we compare the local spectral functions
$A(\omega)$ for both signs of $U$. For positive $U$, as $U$ increases
the upper and lower Hubbard bands emerge and there is a narrow
quasiparticle (QP) band at the Fermi level. For very large $U$, the
low-energy part of the spectrum no longer changes, while the upper
Hubbard band shifts to higher energies \cite{zitko2013}. In the
large-$U$ regime, the system is a doped Mott insulator, which is a
Fermi liquid at low temperatures and a bad metal at high temperatures
\cite{deng}.

For negative $U$, the local spectral function also features Hubbard
bands and a QP peak, but the evolution as a function of $U$ is quite
different. This problem maps onto the half-filled repulsive Hubbard
band in the presence of an external magnetic field of such intensity
that the magnetization remains constant. With increasing $|U|$, the
low-energy scale (Kondo temperature) is reduced exponentially, thus
the QP band shrinks. The negative-$U$ model corresponds to the $B
\sim T_K$ regime in the language of the effective quantum impurity
model with positive $U$. This is precisely the non-trivial cross-over
regime between the well-understood $B=0$ Kondo limit and the
non-interacting $B \to \infty$ limit \cite{hewson2006,bauer2007fm}.
The position of the Hubbard bands is rather symmetric with respect to
zero frequency, but we note the difference in the weight which
corresponds to doping in the $U<0$ picture (or to finite magnetization
in the half-filled effective $U>0$ model picture).

In Fig.~\ref{fig2}(c) we plot the quasiparticle renormalization factor
\begin{equation}
Z(T) = \left( 1-\Re \left[ \frac{d\Sigma(\omega,T)}{d\omega}
\right]_{\omega=0} \right)^{-1}
\end{equation}
at zero-temperature, $Z \equiv Z(T=0)$. It quantifies the renormalized
mass $m^*=m/Z$ and the QP lifetime $\tau^* = Z \tau$.

If the argument of the spectral function is rescaled as $\omega/Z$, we
find that all spectral functions overlap well in the interval $-0.05
\lesssim \omega/Z \lesssim 0.05$. For the positive $U$ case, this
corresponds to the fact that the Fermi liquid regime extends up to
$T_{FL} \approx 0.05 \delta D \approx 0.05 Z D$ (see
Ref.~\onlinecite{deng,zitko2013} and Sec.~\ref{sec5} below).  For the
negative $U$ case, however, this scale ($0.05Z$) is not visible in the
transport properties.

\begin{figure}
\centering
\includegraphics[clip,width=0.48\textwidth]{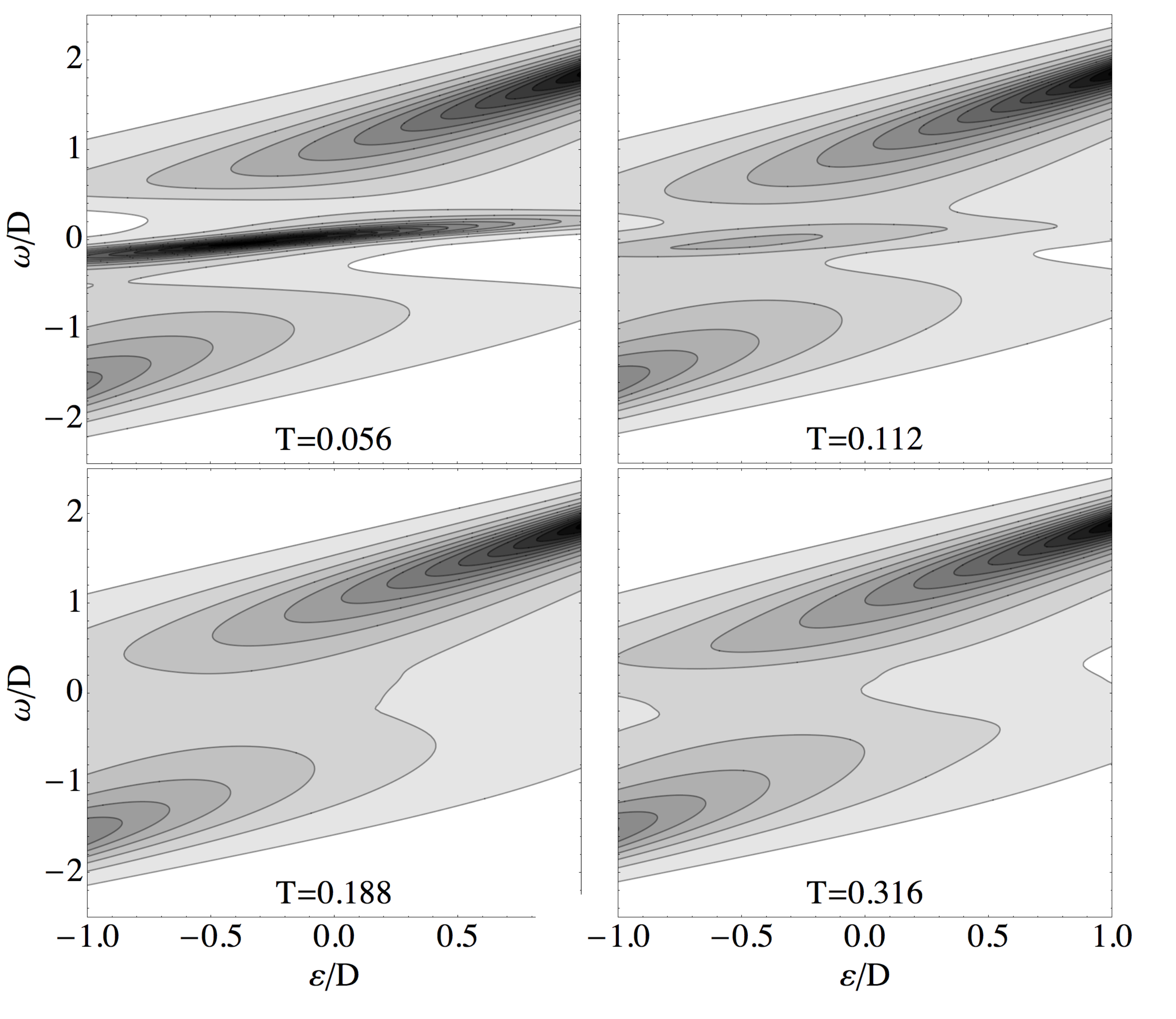}
\caption{(Color online) Momentum-resolved spectral functions
$A(\epsilon, \omega)$ for a range of temperatures for attractive
interaction with $U/D=-2$.}
\label{figk}
\end{figure}

The temperature dependence of spectra for the attractive case is shown
in Fig.~\ref{figk}, where we plot the momentum-resolved
($\epsilon$-dependent) spectral functions. We observe the gradual
disappearance of the QP band (finished by $T \approx 0.15 D$), while
the high-energy Hubbard bands are not affected much in this
temperature range.

\begin{figure}
\centering
\includegraphics[clip,width=0.48\textwidth]{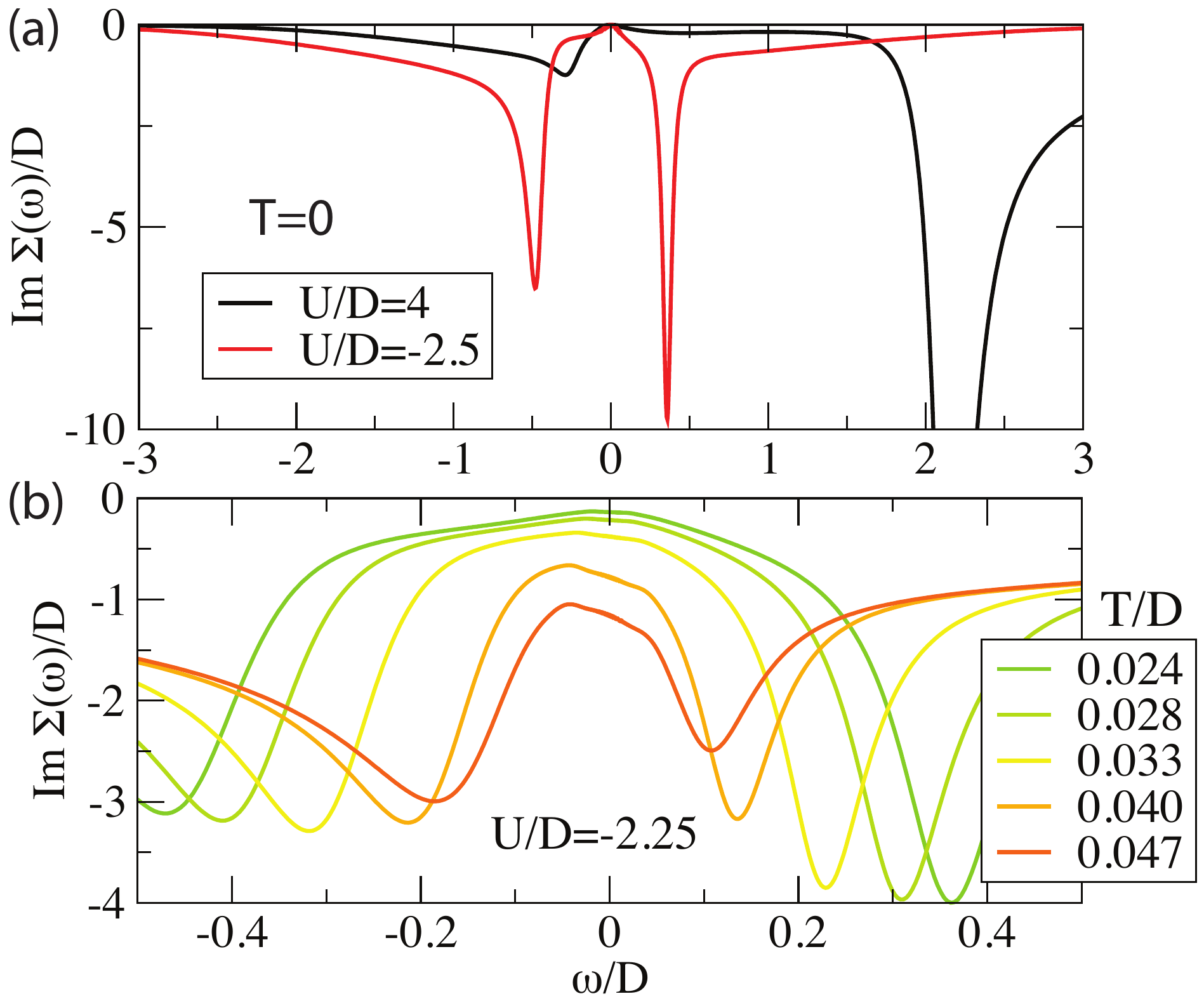}
\caption{(Color online) (a) Imaginary part of the self-energy at $T=0$
for strongly repulsive and strongly attractive interactions reveals
particle-hole asymmetry at low energy scales in both cases. (b)
Temperature dependence of $\Im\Sigma(\omega)$ for the attractive
Hubbard model at $U/D=-2.25$. }
\label{fig3}
\end{figure}

\subsection{Self-energy and particle-hole asymmetry}

We now compare the structure of the self-energy function in repulsive
and attractive case. For weak interaction, they are qualitatively
similar and can be reproduced using the perturbation theory: in $\Im
\Sigma(\omega)$ there are two broad peaks centered approximately at
$\omega=\pm |U|$. For strong interactions, the case shown in
Fig.~\ref{fig3}(a), the differences become more pronounced. The
$U/D=4$ case has been thoroughly studied recently in
Ref.~\onlinecite{deng}, where the strong particle-hole asymmetry in
vicinity of the Fermi level has been pointed out. For strongly
negative $U$ we also find asymmetry in the low-energy part,
but in this case the plateau in $\Im\Sigma(\omega)$ is found on the
{\sl hole side} rather than on the particle side, and it is less
pronounced. In a simplified picture where the asymmetry is related to
the reduced density of states needed for scattering, the long-lived
resilient quasiparticle states for $U>0$ are due to displacement of
the upper Hubbard band to high energies, while the long(er)-lived
quasihole states for $U<0$ are not related so much to the position of
the lower Hubbard band, but rather to its lower spectral weight
(compared with the symmetrically located upper Hubbard band).

For $U<0$, the resonance structures in $\Im\Sigma$ remain rather sharp
on both particle and hole sides; they tend toward small $\omega$ as
$|U|$ increases, which reflects the structure of the spectral function
with shrinking QP band (resonances in $\Im\Sigma$ follow from the
analytical structure of the Green's functions and are expected between
any two spectral peaks in single-orbital problems). For strongly
attractive $U$, the asymmetry decreases for increasing $|U|$.

The temperature dependence of $\Im\Sigma(\omega)$ in the attractive
case reveals an interesting reversal of the asymmetry, see
Fig.~\ref{fig3}(b). This is another non-trivial effect of the
constant-magnetization constraint; it indicates that the $T=0$
self-energy does not permit an easy identification of the transport
mechanisms at elevated temperatures.

\section{Transport properties}
\label{sec5}

In the DMFT, the vertex corrections drop out and the optical
conductivity is fully determined by the self-energy alone
\cite{khurana1990,schweitzer1991,pruschke1993,pruschke1993prb,pruschke1995,jarrell1995,rozenberg1996,georges1996,palsson,bluemer}:
\begin{equation}
\label{sigma}
\Re\, \sigma(\omega) = 
\frac{2\pi e^2}{\hbar} 
\int \mathrm{d}\omega'\, F(\omega,\omega')
\int \mathrm{d}\epsilon\, 
\Phi(\epsilon) 
A_\epsilon(\omega') A_\epsilon(\omega'+\omega),
\end{equation}
with $F(\omega,\omega')=[f(\omega')-f(\omega+\omega')]/\omega$, where
$f(\omega)=(1+\exp(\beta\omega))^{-1}$ is the Fermi function,
$A_\epsilon(\omega)=-(1/\pi)\Im[\omega+\mu-\epsilon-\Sigma(\omega)]^{-1}$,
and $\Phi(\epsilon)$ is the transport function defined through the
derivatives of the dispersion relation:
\begin{equation}
\Phi(\epsilon) = 
\frac{1}{V} \sum_{k} 
\left( 
\frac{\mathrm{d}\epsilon_k}{\mathrm{d}k} 
\right)^2 
\delta(\epsilon-\epsilon_k).
\end{equation}
The expression for $\Re\,\sigma(\omega)$ in Eq.~\eqref{sigma} is valid
generally for a single-band model defined on a lattice which is
periodic and exhibits inversion symmetry in the direction of current
\cite{bluemer}. The Bethe lattice is not a regular lattice and there
is no notion of reciprocal space or momenta, thus there are
ambiguities in the definition of the currents, the optical
conductivity $\sigma(\omega)$, and the transport function
$\Phi(\epsilon)$. We use $\Phi(\epsilon)=\Phi(0)
[1-(\epsilon/D)^2]^{3/2}$, which satisfies the f-sum rule
\cite{chung1998,bluemer,arsenault2013}. The choice of $\Phi(\epsilon)$
has very little effect on the results for the resistivity. It affects
the Seebeck coefficient more significantly, especially for negative
$U$ (where, however, $S$ is small); this is discussed in more detail
in Sec.~\ref{moredet}. In most cases, however, the effects of $\Phi$
are quantitative, not qualitative.

\subsection{Resistivity}

\begin{figure}
\centering
\includegraphics[clip,width=0.48\textwidth]{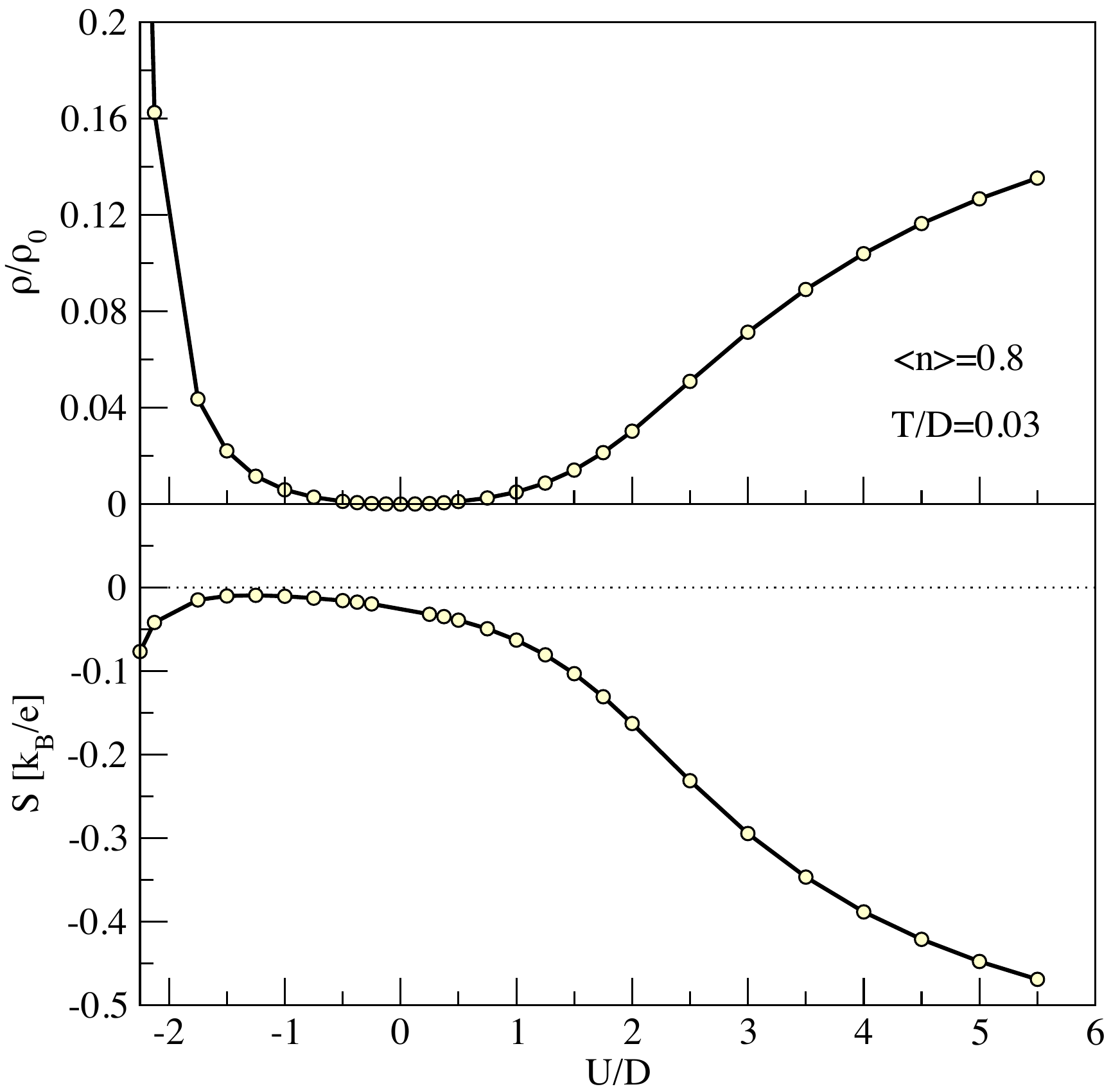}
\caption{(Color online) Resistivity and Seebeck coefficient of the
Hubbard model at constant temperature $T=3\times10^{-2}D$. The
resistivity is expressed in units of the Mott-Ioffe-Regel value
$\rho_0=(e^2/\hbar) \Phi(0)/D$. The results for the Seebeck
coefficient for the values of $U$ where the calculation is not
reliable have been omitted.}
\label{transport}
\end{figure}

We consider first the dc resistivity $\rho=1/\sigma(0)$ at fixed low
temperature as a function of the interaction strength $U$, see
Fig.~\ref{transport}, top panel. The most notable feature is the rapid
resistivity increase for large attraction, $U \lesssim -2D$. This
effect is much stronger than the growing resistivity for increasing
repulsion for $U>0$. This can be explained by the strong decrease of
the effective Kondo temperature, and the corresponding decrease of the
QP lifetime $\tau^*$, see Fig.~\ref{fig2}(c).

\begin{figure}
\centering
\includegraphics[clip,width=0.48\textwidth]{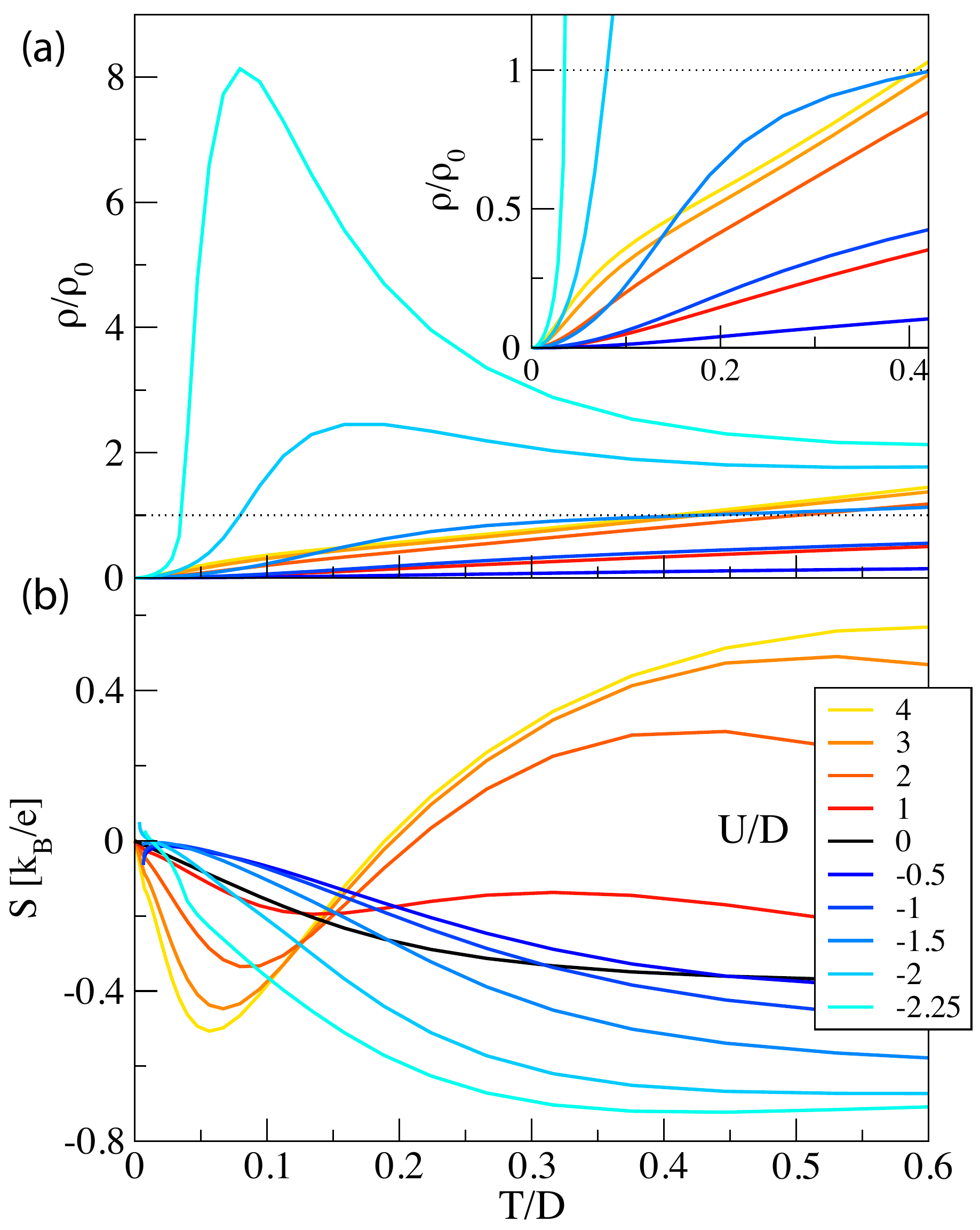}
\caption{(Color online) Temperature dependence of resistivity and
Seebeck coefficient. $\rho_0$ is the MIR resistivity. We note that for
even higher temperatures, not shown in the plot, the resistivity for
$U/D=-2$ and $U/D=-2.25$ starts to increase, i.e., there appears to be
no saturation of resistivity for either sign of $U$. At very low
temperatures (for $T/D \lesssim 0.01$) the results for the Seebeck
coefficient become unreliable due to increasing error in dividing two
small values of the transport integrals $L_{12}$ and $L_{11}$, but
also due to the intrinsic problems of the NRG method in the calculations
of the self-energy function at very small energies and temperatures
(causality violations).
}
\label{transT}
\end{figure}

\begin{figure}
\centering
\includegraphics[clip,width=0.48\textwidth]{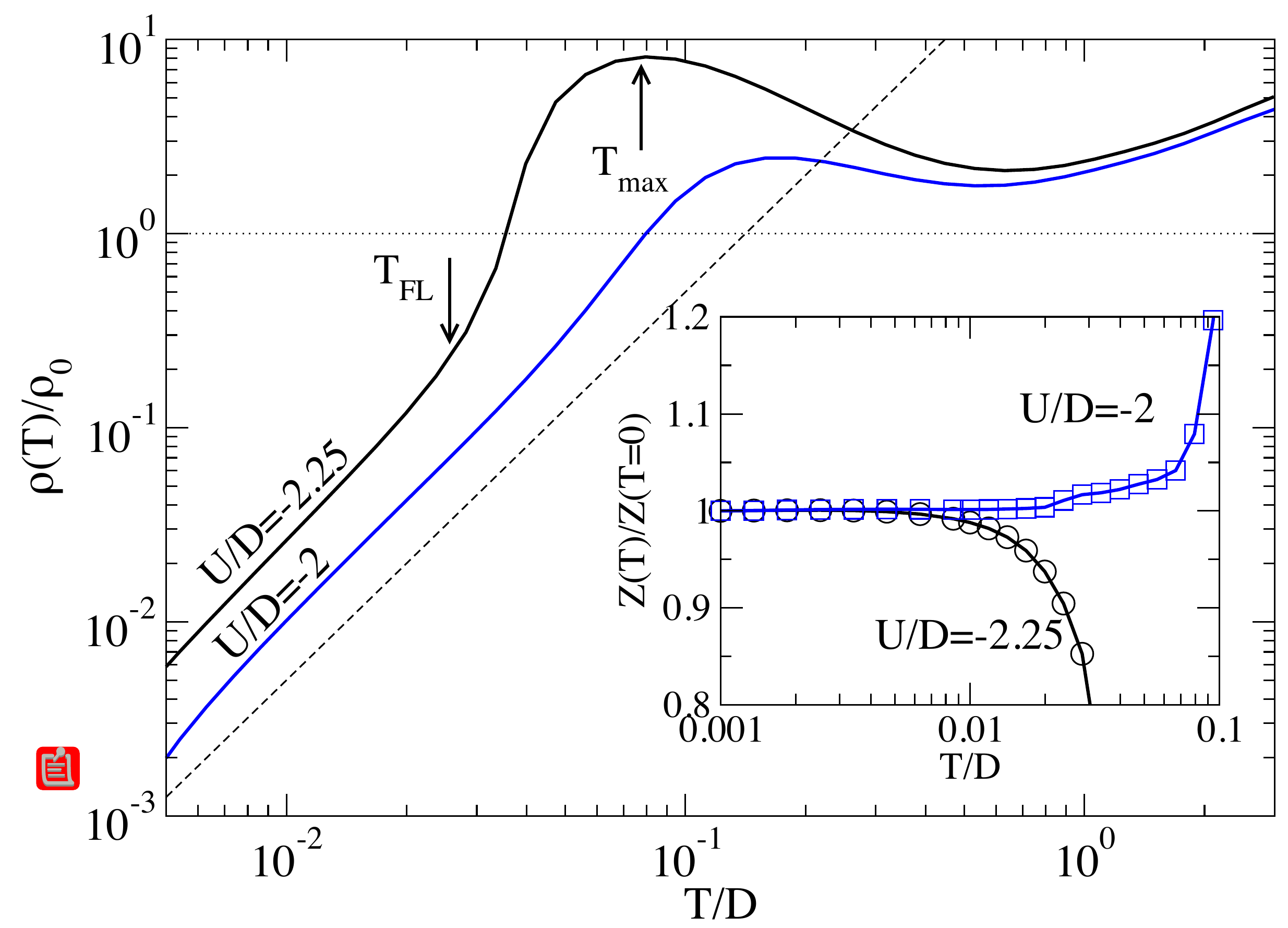}
\caption{(Color online) Resistivity on the log-log scale for
attractive $U$ near the localization transition. The dashed line has
slope 2, expected for the Fermi liquid regime. The dotted horizontal
line indicates the Mott-Ioffe-Regel limit. For $U/D=-2.25$ two
characteristic energy scales can be defined, the Fermi liquid
temperature $T_\mathrm{FL}$ and the resistivity peak temperature
$T_\mathrm{max}$. Inset: rescaled quasiparticle renormalization factor
$Z(T)/Z(T=0)$. Deviation from 1 indicates the end of the Landau Fermi
liquid regime. 
}
\label{rhoLL}
\end{figure}

In Fig.~\ref{transT} we plot the temperature dependence of the
transport properties. At low temperatures, we always find Fermi liquid
behavior $\rho \propto T^2$ below some temperature $T_\mathrm{FL}$ for
$U>U_0$. In the repulsive case, $T_\mathrm{FL}$ is given by
$T_\mathrm{FL} \approx 0.05 \delta D$ where $\delta$ is doping with
respect to half-filling, $\delta=1-\expv{n}$ \cite{deng}. For large
positive $U$, the resistivity above $T_\mathrm{FL}$ increases linearly
with negative intercept up to $T^*$, where the slope changes and the
resistivity is linear with positive intercept \cite{deng}. In the
attractive case, the quadratic dependence extends to much higher
temperatures; for $U/D \gtrsim -2$, it goes essentially up to the
maximum resistivity at approximately $T_\mathrm{max}=ZD$.  For even
stronger attraction, there is a clearer separation between the
$T_\mathrm{FL}$ and $T_\mathrm{max}$ scales, see Fig.~\ref{rhoLL}.
Well-defined QP excitations survive almost up to the high temperature
scale $T_\mathrm{max}$, similar to the resilient quasiparticles
identified in the repulsive case which exist up to $T_\mathrm{MIR}$
where $\rho$ reaches the MIR value \cite{deng}. In the attractive case
at $T_\mathrm{max}$, the resistivity for large enough $|U|$ surpasses
the Mott-Ioffe-Regel limit, thus resilient quasiparticles exist even
in this regime.

\begin{figure}
\centering
\includegraphics[clip,width=0.48\textwidth]{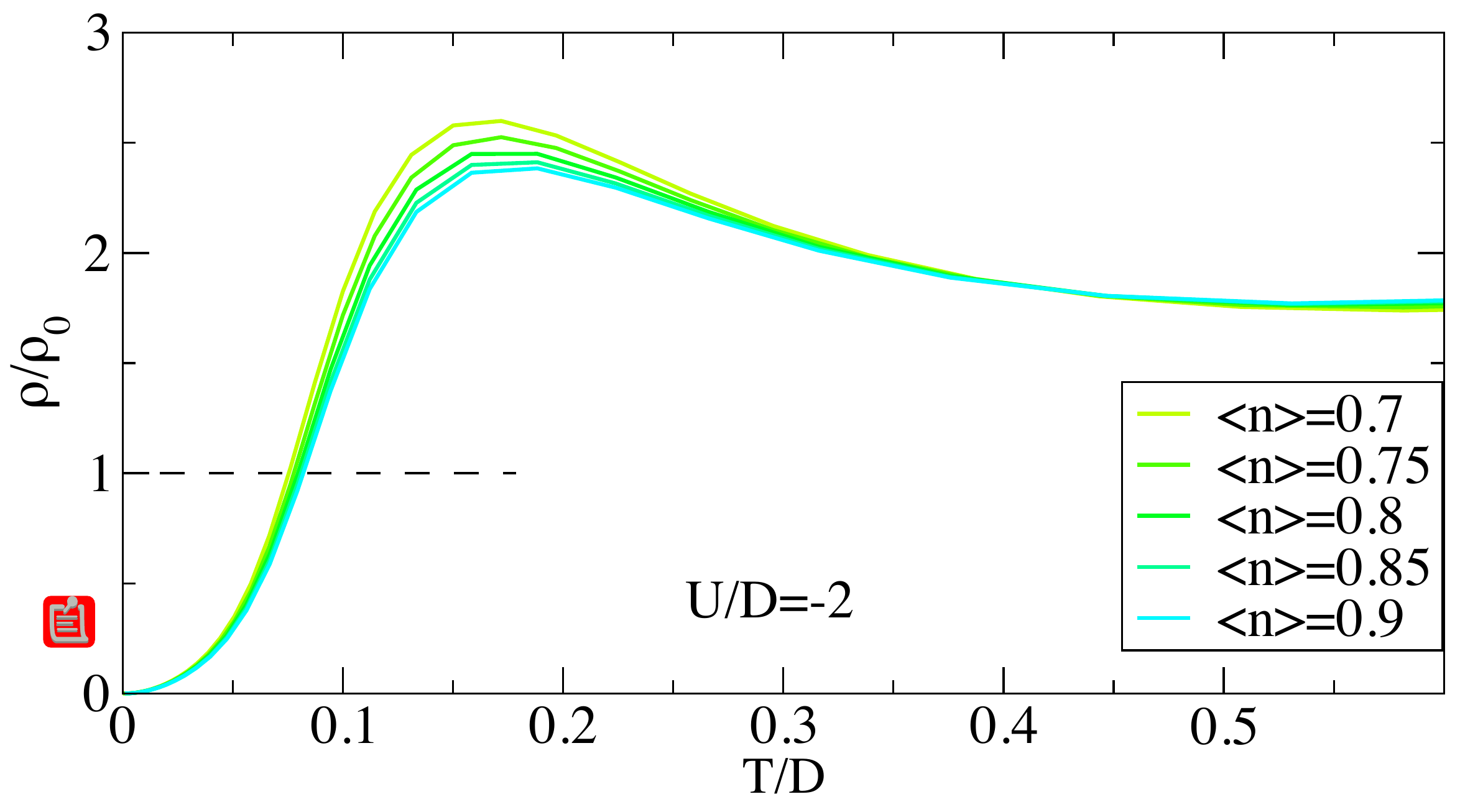}
\caption{(Color online) Resistivity at constant attractive $U$ for a
range of electron densities $\expv{n}$.}
\label{rho}
\end{figure}

While in the repulsive case the characteristic temperature scales
$T_\mathrm{FL}$ and $T_\mathrm{MIR}$ are proportional to doping
$\delta=1-\expv{n}$, in the attractive case the doping does not 
affect much the resistivity curves which are almost overlapping; see
Fig.~\ref{rho}.  $T_\mathrm{max}$ depends mostly on $U$, while the
doping controls the peak value of resistivity, but even this
dependence is found to be very weak. These results can be explained by
the trends seen in the spectral function at low temperature: the QP
band is not affected much by the amount of doping (there is a minor
shift of its low-energy edge, while the high-energy edge is almost
invariant), while there is a significant reorganization of the
spectral weight between the lower and the upper Hubbard band at high
frequencies (this reflects the changing magnetization in the language
of the effective positive-$U$ model at half-filling), but this has
little effect on the resistivity on temperature scales sufficiently
below $\sim |U|/2$.

\subsection{Thermopower (Seebeck coefficient)}

The thermopower (Seebeck coefficient) is defined as
\begin{equation}
S = -\frac{k_B}{e_0 T} \frac{L_{12}}{L_{11}},
\end{equation}
where the transport integrals in the infinite-$d$ limit are
given as \cite{palsson}
\begin{equation}
L_{jk} = \int d\omega\, \left(-
\frac{\partial f(\omega)}{\partial \omega} \right)
\left[ \sum_\sigma \int d\epsilon \Phi(\epsilon)
A_{\sigma,\epsilon}(\omega)^2 \right]^{j}
\omega^{k-1}.
\end{equation}

The results at constant low temperature are shown in
Fig.~\ref{transT}, bottom panel. The Seebeck coefficient for small $U$
is negative because of the asymmetry of the transport function around
the Fermi level (particle-hole asymmetry of electron velocities). For
increasing interaction, it becomes more negative for repulsive $U$ and
less negative for a range of attractive $U$. This behavior can be
explained by the previously discussed asymmetry in the self-energy. On
one hand, the contribution due to the transport function asymmetry is
enhanced due to strong interactions (through a $1/Z$ factor), while
the scattering rate asymmetry depends on the sign of $U$: for $U>0$ it
enhances the absolute value of $S$, while for $U<0$ the two effects
are antagonistic and $|S|$ is reduced. Some further details about the
Seebeck coefficient and the role of the transport function
$\Phi(\epsilon)$ are discussed in Sec.~\ref{moredet}.
The temperature dependence of the thermopower is shown in
Fig.~\ref{transT}(b). For positive $U$, the sign change of $S$ reveals
a change of the dominant transport mechanism and finds its counterpart
in the kink in $\rho(T)$ \cite{deng}. For negative $U$, the Seebeck
coefficient remains negative for all temperatures where reliable
results can be obtained. At very low temperatures it appears to become
positive in a range of temperatures, but those results are
uncertain. Further work with different numerical methods will be
required to clarify the low-temperature behavior of the Seebeck
coefficient in the attractive Hubbard model.

It is interesting to compare these findings for the attractive Hubbard
model with those for the repulsive model at half-filling in the
absence of the magnetic field (zero magnetization) \cite{merino2000}.
The common feature is the non-monotonic behavior of $\rho(T)$ and the
resistivity peak much in excess of the Mott-Ioffe-Regel limit at the
point where the quasiparticles are no longer present. The difference
is found in the behavior of the thermopower.  In the repulsive model,
however, it has a change of sign indicating the thermal destruction of
the coherent Fermi liquid state, similar to what is also found in
doped Mott insulator (i.e., positive $U$ calculations at finite hole
doping, as studied in this work and previously in
Ref.~\onlinecite{deng}).  In the attractive case, the is no such
change of sign.  This qualitative difference in the behavior of
thermopower can be traced back to the partial particle-hole mapping,
Eq.~\eqref{Amap}, and its effect on the transport integrals. $L_{jk}$
includes the factor
\begin{equation}
\label{eq12}
A_{\epsilon,\uparrow}(\omega)^2 + A_{\epsilon,\downarrow}(\omega)^2,
\end{equation}
which maps to
\begin{equation}
\label{eq13}
A_{\epsilon,\uparrow}(\omega)^2 + A_{\epsilon,\downarrow}(-\omega)^2.
\end{equation}
For spin down, the occupied and non-occupied states in the spectral
function are thus interchanged. This mostly affects $L_{12}$ where
the integrand is odd in $\omega$ and thus sensitive to the asymmetry
of spectral functions.

\subsection{Particle-hole asymmetry of the self-energy and the effect 
of different transport functions}
\label{moredet}

We now provide some further details on the dependence of the numerical
results for the transport properties on the choice of the transport
function $\Phi$ \cite{bluemer,arsenault2013}.  Some common
choices are
\begin{itemize}
\item $\Phi_1(\epsilon)=\Phi_0 [1-(\epsilon/D)^2]^{3/2}$,
\item $\Phi_2(\epsilon)=\Phi_0 [1-(\epsilon/D)^2]^{1/2}$, and
\item $\Phi_3(\epsilon)=\Phi_0$.
\end{itemize}

\begin{figure}
\centering
\includegraphics[clip,width=0.48\textwidth]{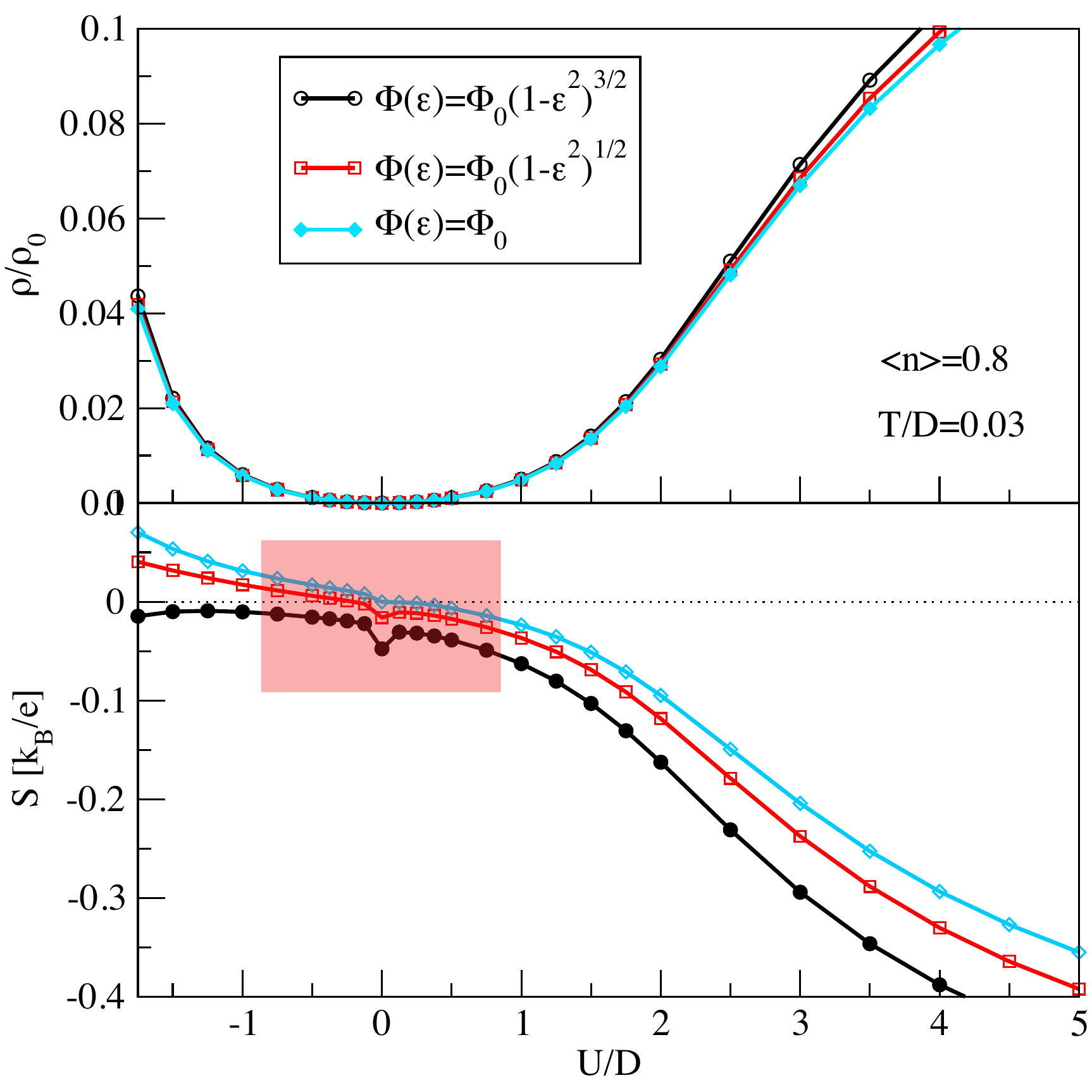}
\caption{(Color online) Resistivity and Seebeck coefficient at
$T/D=0.03$ for three choices of the transport function
$\Phi$.
The data for the Seebeck coefficient in the shaded
rectangle are not reliable for reasons explained in the text.
}
\label{figA1}
\end{figure}

\begin{figure}
\centering
\includegraphics[clip,width=0.48\textwidth]{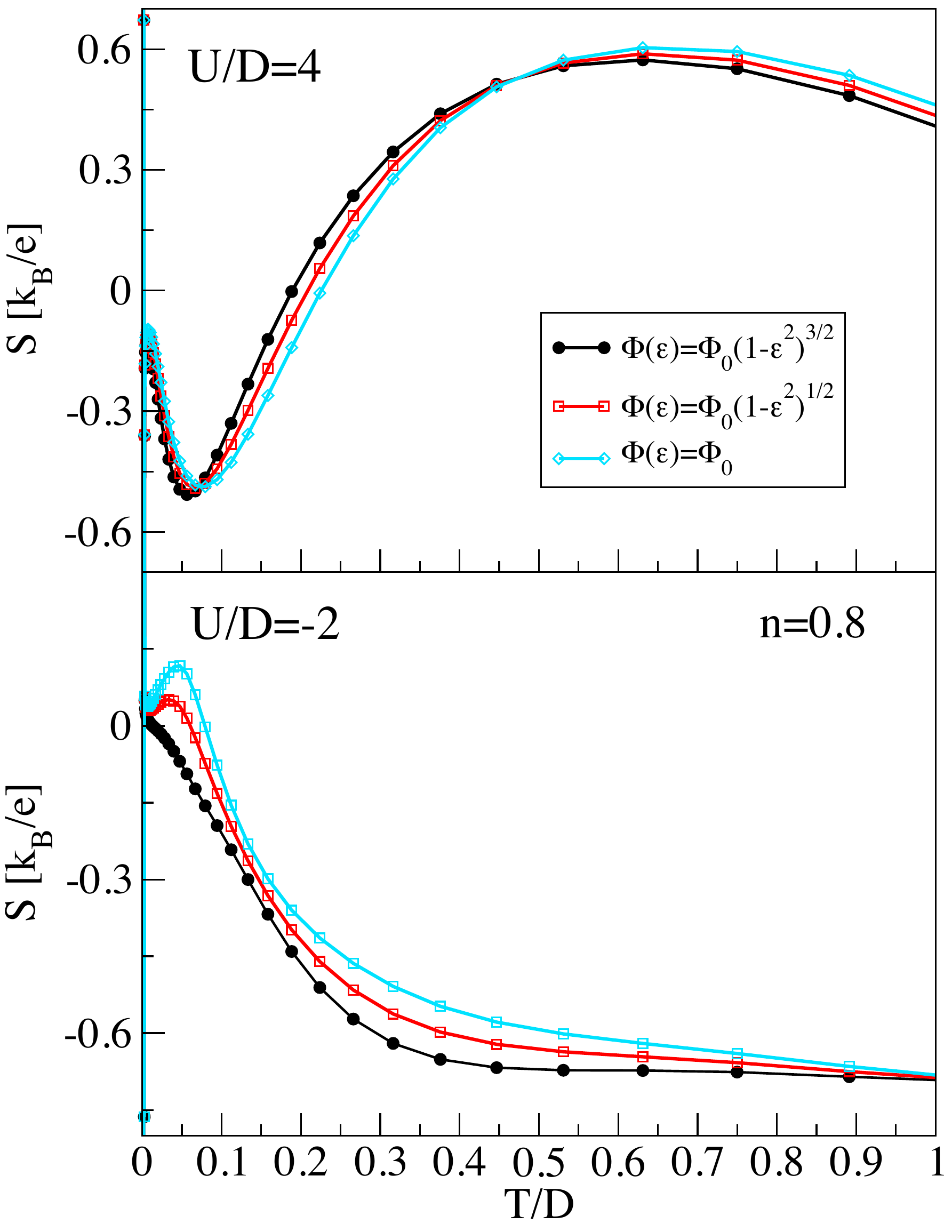}
\caption{(Color online) Seebeck coefficient vs. temperature for two
values of $U$, one positive (top panel) and one negative (bottom
panel), for three choices of the transport function $\Phi$.
The two values are chosen so that they correspond to a comparable value of
the quasiparticle renormalization factor $Z$. The results for $T/D
\lesssim 0.01$ are not correct (note, for example, the downturn of $S$
for $U/D=4$ instead of linear low-temperature behavior).}
\label{figA2}
\end{figure}

In Fig.~\ref{figA1} the results for these three cases are plotted as a
function of $U$ for a fixed temperature $T/D=0.03$. At this moderate
temperature the system is still in the Fermi liquid regime for all
values of $U$ shown in the plot, yet the temperature is sufficiently
high so that the causality-violation issues in the NRG do not affect
the results except for a range of small $U$,
where the oscillatory features in $\Sigma(\omega)$ are not much
smaller than $|\Im\Sigma(\omega)|$ in the relevant frequency interval
$\omega \in [-5T:5T]$ (this is a well known deficiency of the NRG).
In addition, for two values of $U$ we plot the temperature dependence
of the Seebeck coefficient in Fig.~\ref{figA2}.

The resistivity depends little on the choice of $\Phi$, see top panel
in Fig.~\ref{figA1}. In the low-temperature limit, only the value of
$\Phi$ at the Fermi level matters; it enters as a factor in the
Fermi-liquid expression for the resistivity:
\begin{equation}
\rho(T) \propto \frac{1}{Z^2 \Phi(\epsilon_F)} T^2,
\end{equation}
where $\epsilon_F$ corresponds to the Fermi surface value,
$\epsilon_F=\mu-\Re\Sigma(\omega=0)$. Note that $\epsilon_F$ does not
depend on $U$ due to Luttinger's theorem, thus the value of the
prefactor related to $\Phi$ is the same for all $U$.  Even at higher
temperatures beyond the Fermi liquid regime, we find that the
difference is only quantitative. %

The Seebeck coefficient $S$ is more subtle. For repulsive $U$ the
difference is quantitative; when $S$ is plotted as a function of the
temperature, the effect of different $\Phi$ is mainly a 
slight shift of the characteristic temperatures (positions of extrema
and zero-crossings), but it hardly affects the overall scale of $S$
(in particular the values at the minima and maxima); see
Fig.~\ref{figA2}, top panel. This is not the case of attractive $U$,
where we observe {\sl qualitatively different} behavior at low
temperatures: the sign itself of the low-temperature slope of the
Seebeck coefficient depends on the choice of $\Phi$; see
Fig.~\ref{figA2}, bottom panel. 

As first pointed out by Haule and Kotliar \cite{haule2009}, the
particle-hole asymmetry terms in the low-frequency expansion of
$\Im\Sigma$ may change the slope of $S(T)$ compared to the
Fermi-liquid estimate which retains only the lowest order $\omega^2 +
(\pi k_B T)^2$ terms \cite{deng}. The full expression for the Seebeck
coefficient in the low-temperature limit is \cite{haule2009}
\begin{equation}
\label{SS}
S=- \frac{k_B}{e_0} \frac{k_B T}{Z}
\left( \frac{E_2^1}{E_0^1}
\frac{\Phi'(\epsilon_F)}{\Phi(\epsilon_F)}
- 
\frac{a_1 E^2_4 + a_2 E^2_2}{\pi^2 \gamma_0 E^1_0} 
\right),
\end{equation}
where $E^k_n$ are numerical constants of order unity defined as
\begin{equation}
E^k_n = \int \mathrm{d}x \frac{x^n}{\cosh(x/2)^2}
\frac{1}{(1+x^2/\pi^2)^k},
\end{equation}
$a_i$ are the expansion coefficients of the cubic self-energy terms:
\begin{equation}
\Sigma^{(3)}(\omega) = \frac{a_1 \omega^3+a_2 \omega T^2}{Z^3},
\end{equation}
and $\gamma_0$ is defined as the prefactor of the quadratic terms:
\begin{equation}
\Sigma^{(2)}(\omega) = \frac{\gamma_0}{Z^2} \left( \omega^2+\pi^2 k_B^2
T^2 \right).
\end{equation}
The first term in Eq.~\eqref{SS} describes the particle-hole asymmetry
in the electronic velocities, the second the asymmetry in the
scattering rate. %
For fixed $n$, $\Phi'(\epsilon_F)/\Phi(\epsilon_F)$ is a fixed value
that depends only on the choice of the function $\Phi$. It is zero for
$\Phi(\epsilon)=\text{const}$, and it differs by a factor of three for
$\Phi(\epsilon)=\Phi_0 (1-\epsilon^2)^{1/2}$, where
\begin{equation}
\frac{\Phi'(\epsilon_F)}{\Phi(\epsilon_F)} = -\frac{\epsilon_F}{1-\epsilon_F^2},
\end{equation}
and $\Phi(\epsilon)=\Phi_0 (1-\epsilon^2)^{3/2}$, where
\begin{equation}
\frac{\Phi'(\epsilon_F)}{\Phi(\epsilon_F)} = -
\frac{3\epsilon_F}{1-\epsilon_F^2}.
\end{equation}
For different choices of $\Phi$, the contribution of the
first term in Eq.~\eqref{SS} forms a progression $0, 1, 3$, which
thus forms a gauge to assess its importance compared to the second
term.

In Ref.~\onlinecite{deng}, it was shown that at repulsive $U/D=4$ the
particle-hole asymmetry in the self-energy leads to a change of slope
by a factor of more than 2. The rather small dependence of the slope
of $S(T)$ on the choice of $\Phi$ seen in Fig.~\ref{figA2}, top panel,
actually suggests that the particle-hole asymmetry of
$\Im\Sigma(\omega)$ is the {\sl dominant contribution to the
thermopower} for large $U$.

For attractive $U/D=-2$, the situation is even more interesting. Due
to the asymmetry with long-lived hole states (see Fig.~\ref{fig3}),
the second term in Eq.~\eqref{SS} has a different sign from the first
one. Since, in addition, the two terms are of similar magnitude, even
the {\sl sign of the slope} is affected by the choice of
$\Phi$.

Of course, for a real lattice the transport function $\Phi$ is fully
determined by the dispersion relation and there is no element of
indeterminacy. Nevertheless, the foregoing analysis has shown that the
asymmetry term can be as large as or even larger than the first
lowest-order Fermi-liquid term, possibly reversing the sign of the
Seebeck coefficient. Proper inclusion of corrections to the Fermi
liquid theory are thus crucially (i.e., qualitatively) important for
hole-doped systems with long-lived resilient quasihole states and
electron-doped ones with long-lived quasiparticle states, and
quantitatively important in general.

\section{Optical conductivity}

\begin{figure}
\centering
\includegraphics[clip,width=0.48\textwidth]{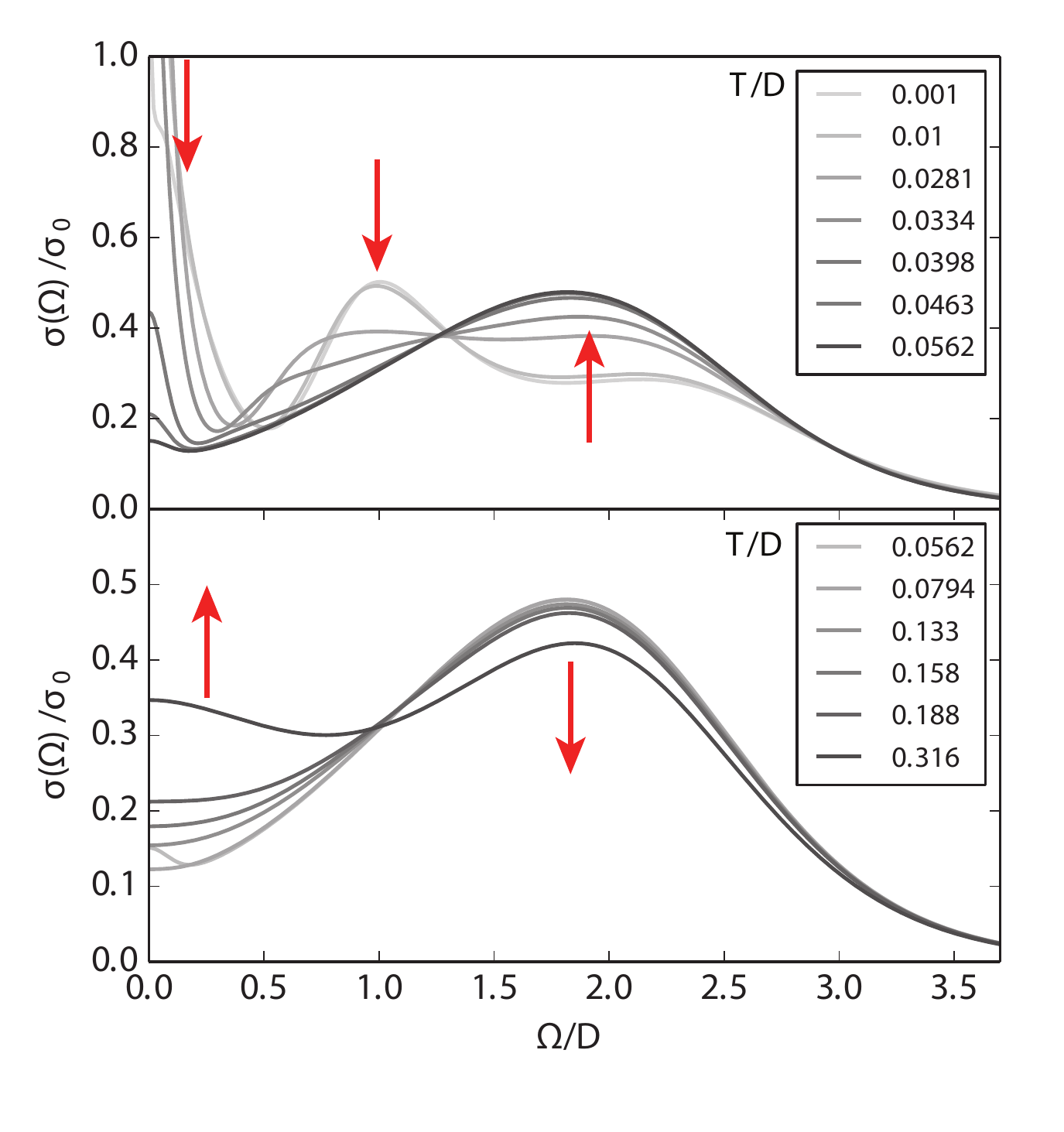}
\caption{(Color online) Optical conductivity for attractive
$U/D=-2.25$ for a range of temperatures. The arrows indicate the
evolution with increasing temperature in different frequency regions.
Upper panel roughly corresponds to $T \lesssim T_\mathrm{max}$,
lower to $T \gtrsim T_\mathrm{max}$.
} 
\label{opt}
\end{figure}

For both signs of $U$, the optical conductivity at low temperatures
shows the well-known characteristics of the Fermi-liquid state in the
Hubbard model \cite{pruschke1993prb,georges1996}: a pronounced Drude
peak at $\Omega=0$ due to transitions inside the QP band, peak(s) or a
band corresponding to transition between the QP band and the Hubbard
bands near $\Omega=|U|/2$ or up to $\Omega \sim D$ (mid-infrared
region), and a more diffuse peak at $\Omega=|U|$ due to the
inter-Hubbard-band excitations. The results for attractive interaction
$U/D=-2.25$ are shown in Fig.~\ref{opt}. At low temperatures the peaks
are rather well defined and clearly separated. As the temperature
increases, the Drude peak intensity decreases. For $T \gtrsim
T_\mathrm{FL}$, the intensity of the peak at $\Omega \approx |U|/2$
also drops and shifts toward lower frequencies. In this temperature
range of $T \lesssim T_\mathrm{max}$, the optical spectral weight is
transferred mostly to the $\Omega=|U|$ inter-Hubbard-band peak. As the
temperature is increased further to $T \gtrsim T_\mathrm{max}$, there
is a spectral redistribution in the opposite direction, from the
$\Omega=|U|$ region to low-frequency regions, which corresponds to the
decreasing dc resistivity in the temperature interval from
$T_\mathrm{max}$ to the plateau of nearly constant resistivity around
$T = 0.5 D$, as seen in Figs.~\ref{transport} and \ref{transT}. (For
repulsive case, the temperature dependence of $\sigma$ was studied in
Ref.~\onlinecite{deng}.)

\begin{figure*}
\centering
\includegraphics[clip,width=0.98\textwidth]{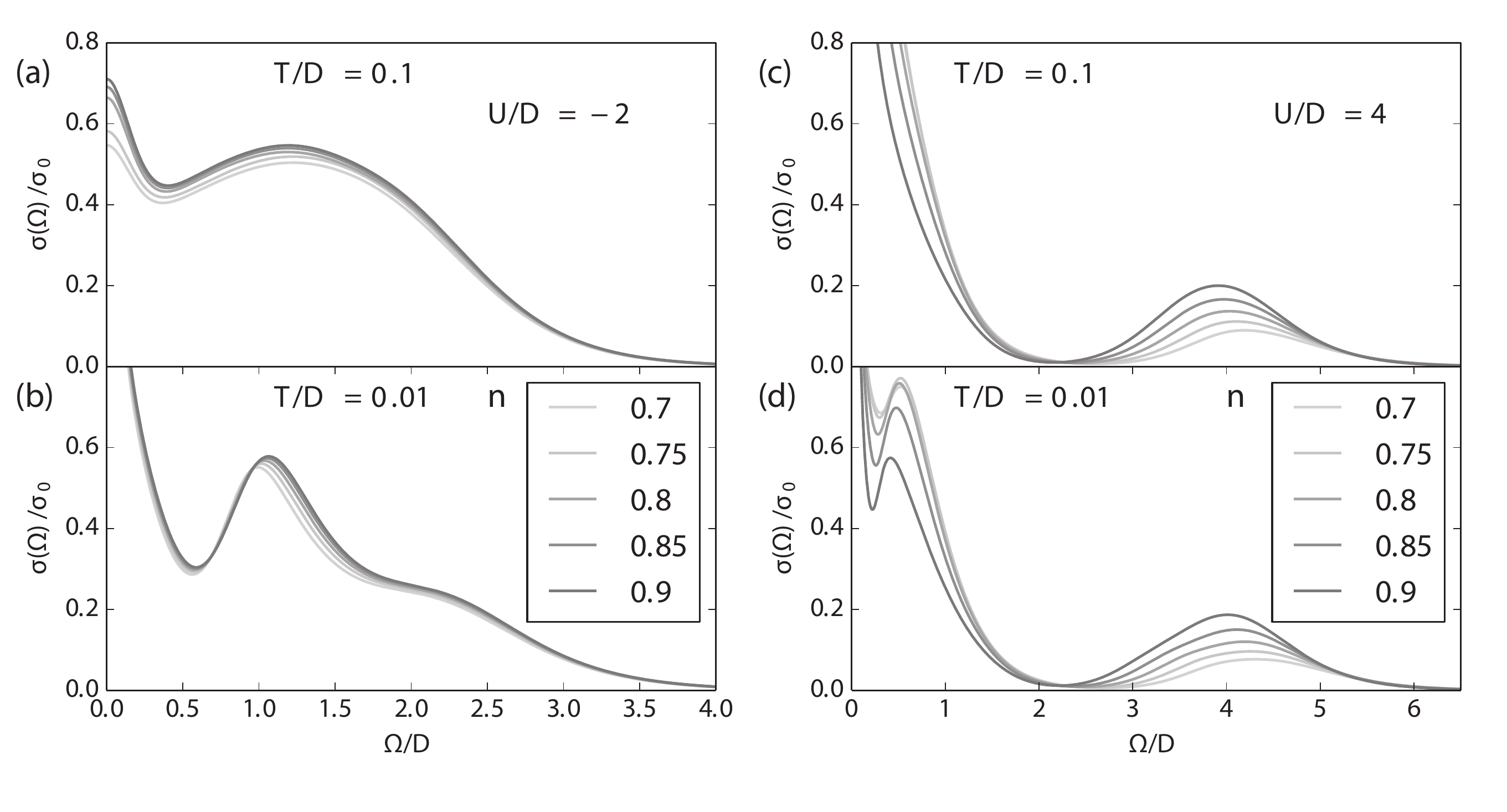}
\caption{(Color online) Optical conductivity for $U/D=-2$ (left) and
$U/D=4$ (right) for a range of band fillings $n$.}
\label{figE}
\end{figure*}

For completeness, we also study the $n$-dependence of the optical
conductivity at two characteristic temperature regimes ($T/D=10^{-2}$ is
well in the Fermi liquid regime, $T/D=10^{-1}$ corresponds to the
cross-over regime between the low-temperature and high-temperature
asymptotics) for both signs of $U$, see Fig.~\ref{figE}. 

For positive $U$, the results for the lower temperature $T/D=10^{-2}$
(bottom right panel in Fig.~\ref{figE}) are easy to understand. With
increasing doping (decreasing $n$), both Hubbard bands shift to higher
energies, thus the corresponding optical peaks also move up. At the
same time, the spectral weight of the QP band is increasing, while
that of the Hubbard bands is decreasing; the system is becoming less
correlated. This is reflected in the decreasing weight of the optical
peak at $\Omega \approx U$ (upper Hubbard band, UHB), although that at
$\Omega \approx 0.5 D$ (lower Hubbard band, LHB) is actually
increasing due to the increasing density of initial QP states. At
higher temperature $T/D=10^{-1}$ (upper right panel in
Fig.~\ref{figE}), the QP-LHB transitions can no longer be resolved,
but the general trend with increasing doping is similar as in the
Fermi-liquid regime.

For strong attraction, the optical conductivity is expected to weakly
depend on doping, since in the effective model the changing
magnetization leads to a rather moderate spectral weight
redistribution: it mostly affects the total weight in the atomic
peaks, while their positions remain essentially the same. The results
are in agreement with the trends in the dc resistivity, shown in
Fig.~\ref{rho}. The most significant variation of the dc resistivity
is found in the peak region from $T \approx 0.1D$ to $T \approx 0.2
D$: in this temperature range the optical conductivity is affected on
an extended frequency range from $\Omega=0$ up to $\Omega \approx 2D$
which includes the transitions inside the QP band and between the QP
band and either Hubbard band: the main effect is that with increasing
doping the optical conductivity decreases almost uniformly, with no
changes in peak positions (upper left panel in Fig.~\ref{figE}). The
behavior is different at the lower temperature of $T=0.01D$ (bottom
left panel in Fig.~\ref{figE}): the main effect there is a shift in
the upper flank of the peak in $\sigma(\Omega)$ at $\Omega \approx
|U|/2$, which corresponds to the transitions between the QP band and
either Hubbard band, but little overall decrease in the optical
conductivity.

\section{Spin-lattice relaxation rate and dynamical susceptibilities}
\label{sec6}

The spin susceptibility can be probed in nuclear magnetic
resonance (NMR) experiments. The spin-lattice relaxation rate $1/T_1$
quantifies the decay of the nuclear magnetic moments and provides
information about the fluctuations of the electronic magnetic moments:
\begin{equation}
\frac{1}{T_1} = 2k_B T \left( \frac{g_N \mu_N}{g\mu_B} \right)^2
\sum_q |H_\mathrm{hf}(q)|^2 \Im\left[
\frac{\chi^{+-}(q,\omega_N)}{\omega_N} \right],
\end{equation}
where $\omega_N$ is the nuclear Larmor frequency which may be set to
zero. If the hyperfine interaction $H_\mathrm{hf}(q)$ is local (i.e.,
has very weak $q$ dependence), we are effectively probing the local
dynamical magnetic susceptibility that is easily computed using the
NRG. Furthermore, if there is no magnetic order,
$\chi_{zz}=\frac{1}{2}\chi_{+-}$ due to isotropy in spin space. Thus,
in the context of paramagnetic DMFT calculations, $1/T_1T$ measures
the slope of the imaginary part of $\chi_\mathrm{loc}$ in the
zero-frequency limit.

\begin{figure*}
\centering
\includegraphics[clip,width=0.98\textwidth]{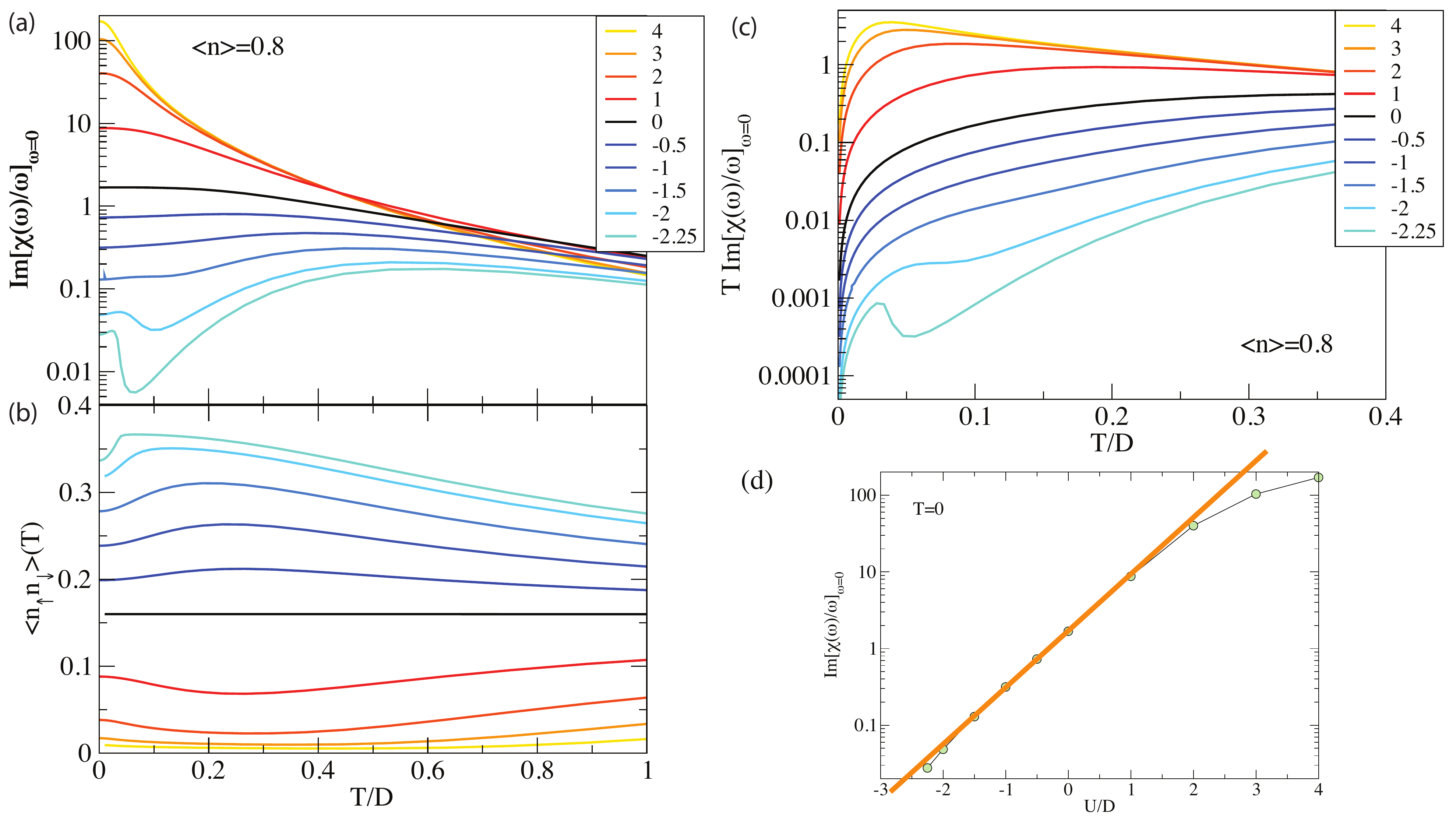}
\caption{(Color online) (a) Zero-frequency slope of the imaginary part
of the local dynamical spin susceptibility, i.e., $1/T_1T$ up to
constant prefactor. (b) Double occupancy as a function of temperature.
(c) Zero-frequency slope multiplied by the temperature, i.e., $1/T_1$.
(d) Zero-temperature spin relaxation rate vs. Hubbard parameter $U$. }
\label{NMRT}
\end{figure*}

The temperature dependence of the relaxation rate is shown in
Fig.~\ref{NMRT}(a), where we plot the zero-frequency slope of the
dynamical magnetic susceptibility (i.e., $1/T_1$), and
Fig.~\ref{NMRT}(c), where this same quantity is multiplied by the
temperature (i.e., $1/T_1T$). For strongly repulsive interaction, the
relaxation rate $1/T_1$ is monotonously decreasing with temperature:
for $U/D=4$ it drops by four orders of magnitude when going from $T=0$
to $T \sim D$. For attractive $U$, the dependence is more complex and
non-monotonous. The case of $U/D=-2$ is typical for the strongly
attractive regime. The pronounced minimum at $T\sim 0.1D$ corresponds
to the maximum in $P_2(T)=\expv{n_\uparrow n_\downarrow}(T)$, see
Fig.~\ref{NMRT}(b): higher double occupancy (pairing) implies less
developed local moments. In the repulsive case the behavior is
opposite: $P_2$ starts by decreasing upon heating leading. In both
cases this leads to increased localization, which can be explained by
the higher entropy in the Mott insulating (respectively pairing) phase
\cite{georges1996}. We also generally observe that the scale of
temperature variations is significantly smaller in the $U<0$ case as
compared to the $U>0$ case. The presentation of the results as
$1/T_1T$ in Fig.~\ref{NMRT}(c) indicates the low-temperature metallic
behavior (proportional to $T$) and regions of insulator-like behavior
with nearly constant $1/T_1$ (in particular the bad-metal regime for
large repulsive $U$).

The relaxation rate at $T=0$ is plotted in Fig.~\ref{NMRT}(d). The
general trend is expected: for the repulsive $U$ the system exhibits
sizeable magnetic fluctuations which saturate in the large-$U$ limit,
while for the attractive $U$ the spin fluctuations rapidly freeze out. In
the interval $-D < U < D$, $1/T_1T$ depends exponentially on $U$,
approximately as
\begin{equation}
\frac{1}{T_1T} \propto \exp\left(d \frac{U}{D}\right),
\quad\text{with}\quad d \approx 1.7.
\end{equation}
For a more strongly attractive $U$, the reduction becomes even more
pronounced. This is associated with the emergence of the sharp Kondo
resonance in the charge sector, while the spin fluctuations become
negligible.

\begin{figure}
\centering
\includegraphics[clip,width=0.5\textwidth]{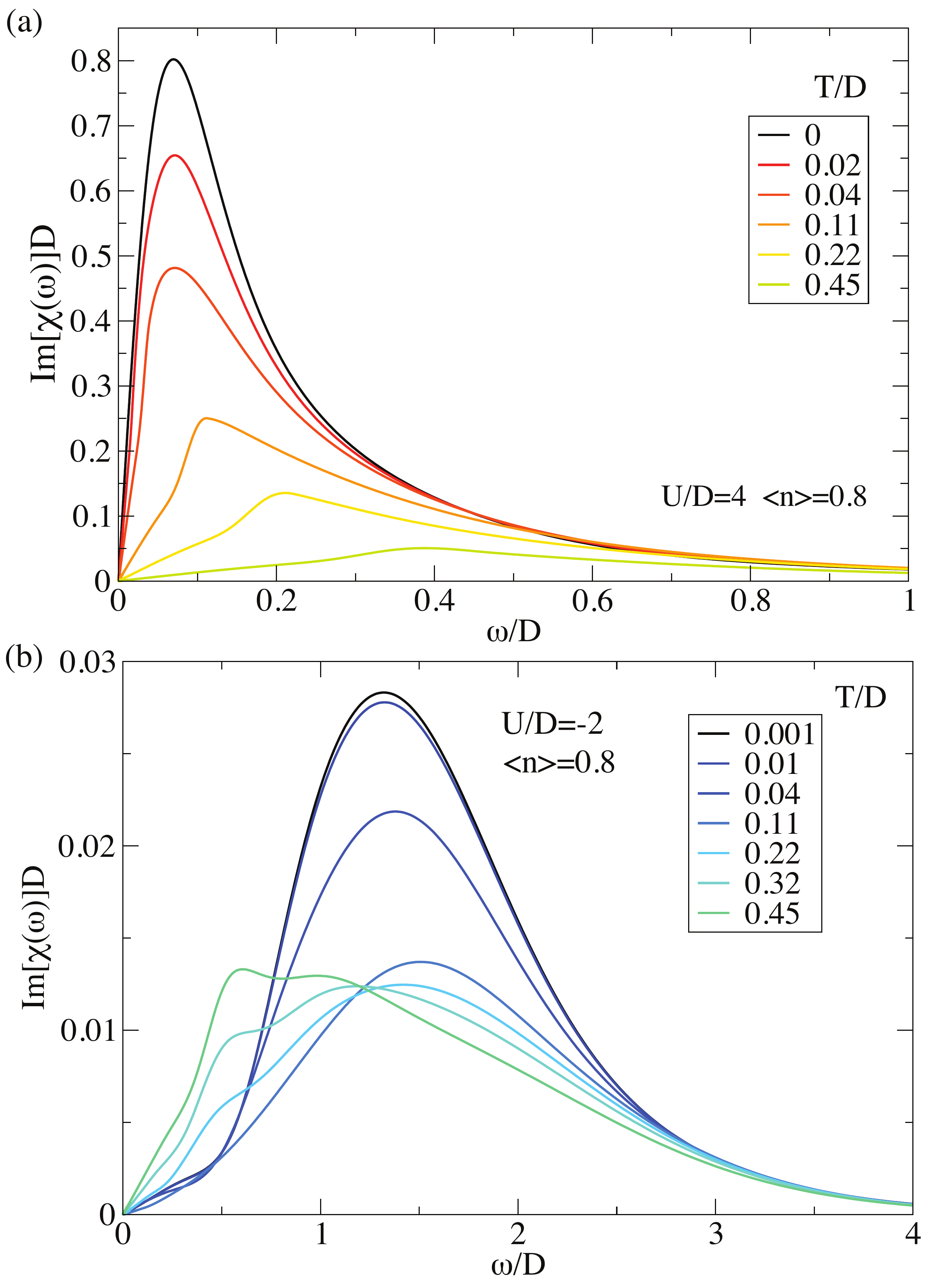}
\caption{(Color online) Imaginary part of the dynamical spin
susceptibility for a range of temperatures for (a) repulsive
interaction and (b) attractive interaction.}
\label{imchi}
\end{figure}

In Fig.~\ref{imchi} we show local dynamical spin susceptibility
for a range of temperatures, one set for a representative
case of repulsive (top panel) and one for attractive interaction
(bottom panel). 
For $U/D=4$, the dominant peak is on the Kondo scale with a maximum
close to $\omega_\mathrm{sf} \approx 0.3ZD \approx 0.07$; this
corresponds to the coherence scale of the problem \cite{georges1996}.
This peak corresponds to the fluctuations of the local moments which
is screened in the lattice version of the Kondo effect and is
generated by the particle-hole excitations in the quasipartile band. A
much weaker peak (off scale in the plot) is present on the scale of
charge fluctuations at $\omega \approx U$ due to particle-hole
excitations with the hole in the LHB and the particle in the UHB.
For temperatures below $T_\mathrm{coh} \approx \omega_\mathrm{sf}$ the
susceptibility peak maximum remains close to $\omega_\mathrm{sf}$,
only its amplitude is decreasing with increasing temperature.
For $T \gtrsim T_\mathrm{coh}$ the peak maximum itself shifts to
higher frequencies; in fact, in this temperature regime the maximum
occurs at $\omega \approx T$. 

\begin{figure}
\centering
\includegraphics[clip,width=0.5\textwidth]{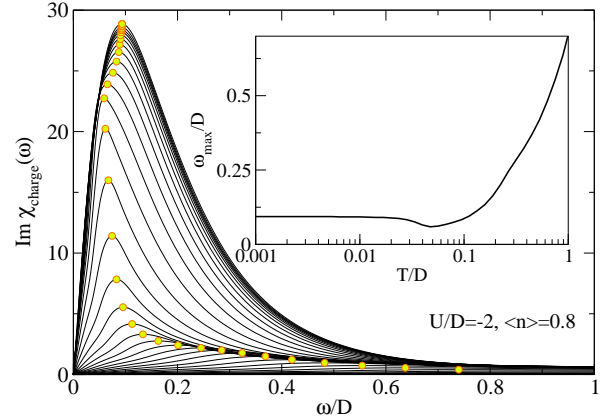}
\caption{(Color online) Dynamical charge susceptibility of the
attractive Hubbard model for a range of temperatures. The insets shows
the frequency of the peak as a function of the temperature.}
\label{fignn}
\end{figure}

In the repulsive case, the spin fluctuations are expectedly much
weaker. At $T=0$ there is a single peak on the scale of $\omega \sim
D$ and some non trivial structure on the low-frequency scale of $\sim
Z D$.  The temperature variation is quite complex. Regime 1: Up to
$T/D \sim 0.04$, the main effect is some reduction of weight in the
high frequency region, while the low frequency region that determines
$1/T_1$ is largely unaffected. Regime 2: For $T/D$ between $\approx
0.04$ and $\approx 0.11$, there is a reduction of spin fluctuations on
all energy scales, which corresponds to decreasing $1/T_1T$. Regime 3:
For $T/D>0.11$ a new peak starts to develop in the low-frequency
region, while the high-energy peak shifts to lower frequencies; the
two peaks merge at very high temperatures of order bandwidth. The
crossovers between the regimes find their counterparts in the
temperature dependence of the entropy (see Fig.~\ref{fig1b}). The
crossover between regimes 1 and 2 corresponds to the emergence of an
entropy plateau due to increasing pairing between the electrons. 
These pairs would condense into a coherent superconducting state if
superconducting order were allowed in our calculations. This crossover
is not visible, however, in the transport properties: the resistivity
is almost perfectly quadratic in both regimes 1 and 2 with no visible
kinks, see Fig.~\ref{rhoLL}. The crossover between regimes 2 and 3 can
be interpreted as a thermal decomposition of the electron pairs. These
regimes can also be observed in the dynamical charge susceptibility
shown in Fig.~\ref{fignn}. For $U<0$ this quantity behaves somewhat
similarly as the dynamical spin susceptibility for $U>0$, except for a
softening of the charge fluctuation mode in the temperature range
between regimes 1 and 2 (the position of the peak vs. $T$ is shown in
the inset).

The fine details in the dynamical susceptibility curves for $\omega
\lesssim T$ should be interpreted with care due to possible artifacts
\cite{pade}. In this respect, two-particle properties are even more
challenging to determine reliably in the NRG at finite $T$ than the
single-particle properties. In particular, it is difficult to answer
the question if a zero-frequency $\delta$ peak is present in the
greater Green's function $\Im \chi^>(\omega)$ at $T>0$ as might be
expected for unscreened local moments in the bad metal regime of a
doped Mott insulator. We indeed observe a $\delta$ peak develop as $T$
is increased, but it can be shown that due to the particular way the
spectra are computed in the NRG, some part of its weight is likely to
be unphysical (see Appendix \ref{appA}). Unfortunately, it is unclear
how to separate the two contributions. In spite of these difficulties,
the spin-lattice relaxation rate $1/T_1$ can be extracted relatively
robustly from the retarded Green's function $\Im \chi(\omega)$ after
spectral broadening with a kernel of width $\sim T$ and performing a
linear fit in an interval of width $\Delta \omega \sim T$ around
$\omega=0$.

\section{Discussion: Anderson impurity at constant magnetization}
\label{sec7}

\begin{figure*}
\centering
\includegraphics[clip,width=\textwidth]{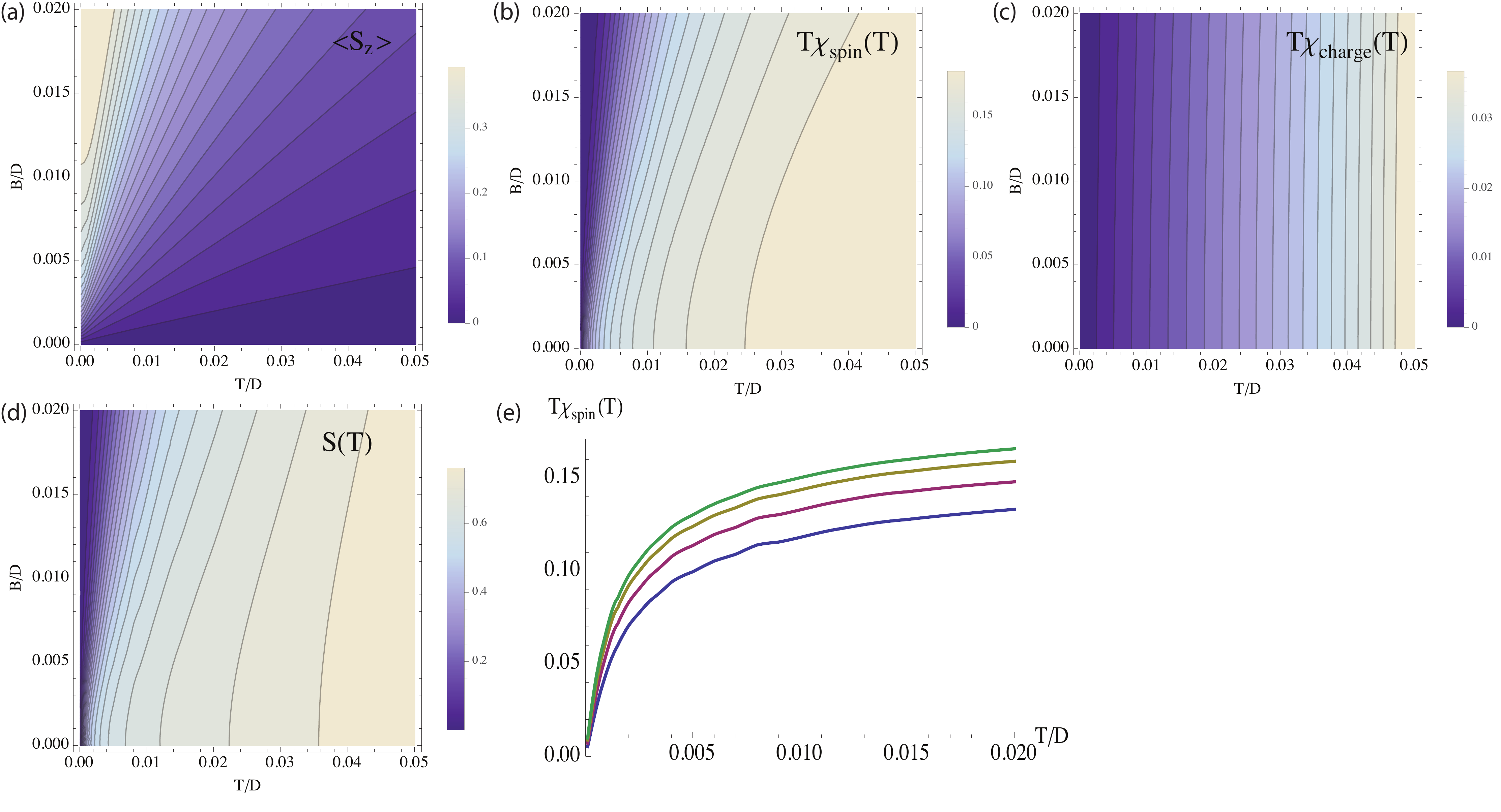}
\caption{(Color online) Properties of the single impurity Anderson
model at half-filling as a function of the temperature $T$ and the
magnetic field $B$. (a) Magnetization, (b) impurity spin
susceptibility, (c) impurity charge susceptibility, and (d) impurity
entropy. (e) Temperature dependence of the impurity spin
susceptibility evaluated along the constant-magnetization contours
(top to bottom: $\expv{S_z}=0.05, 0.1, 0.1, 0.2$). }
\label{siam}
\end{figure*}

\begin{figure}
\centering
\includegraphics[clip,width=0.48\textwidth]{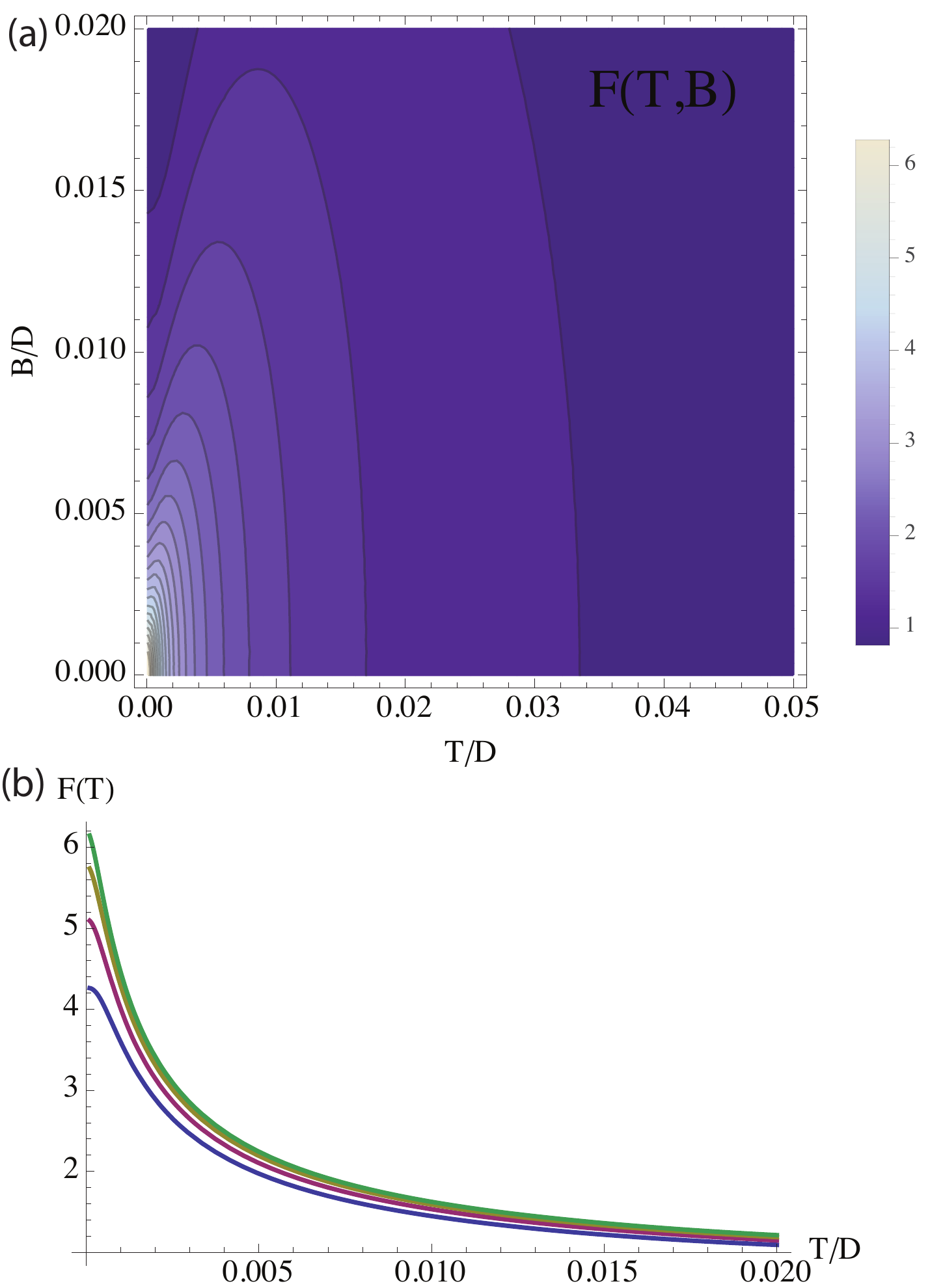}
\caption{(Color online) Conductivity $F(T,B)$ for a single spin
species in the single-impurity Anderson model at half-filling in
external magnetic field (see the text for the exact definition). (a)
Contour plot in the $(T,B)$ plane. (b) Temperature dependence of the
conductivity along the constant-magnetization contours (top to bottom:
$\expv{S_z}=0.05, 0.1, 0.15, 0.2$). }
\label{siamV2}
\end{figure}

The non-monotonic temperature dependences in the attractive-$U$
Hubbard model have been explained through the non-trivial properties
of the positive-$U$ model in magnetic field at constant magnetization.
In this section, we investigate to what extent this behavior is
present already at the level of the quantum impurity model without the
self-consistency loop. In other words, we consider the Anderson
impurity model at the particle-hole symmetric point as a function of
the external magnetic field $B$ and the temperature $T$, and study its
properties along the constant magnetization contours. The magnetic
field is expressed in energy units (i.e., it includes the
$g\mu_B$ prefactor, where $g$ is the $g$-factor and $\mu_B$ the Bohr
magneton). We choose $U/D=0.5$, $\delta=0$ and a flat band with
constant hybridization function $\Gamma/D=0.05$. For this parameter
set, the Kondo temperature according to Wilson's definition is
$T_K/D=10^{-3}$. We consider a temperature range up to $T=0.05D = 50
T_K$, where the Kondo peak is already strongly suppressed (but {\sl
still visible as a small hump} at the Fermi level), and magnetic
fields up to $B=0.02D=20 T_K$, where the spin polarization at low
temperatures is $80\%$ and there is a strong Kondo peak splitting
(although the Zeeman-split peaks are {\sl still clearly present}). The
persistence of non-trivial low-frequency spectral features at $T$ and
$B$ of severals tens of $T_K$ are worth stressing again: the Kondo
effect is a cross-over with logarithmic dependencies, thus it affects
the system properties in a {\sl wide temperature and field range much
above the $T_K$ scale}. This has obvious implications for the physics
of the Hubbard model considered within the DMFT approach, since a
quasiparticle band must consequently be present on temperature scales
much above $Z \sim T_K$, unless suppressed through the additional
effect of the DMFT self-consistency constraint.

In Fig.~\ref{siam}(a) we plot the constant-magnetization contours in
the $(T,B)$ plane. For low magnetization, the contours are almost
linear: curvature is visible only at low temperatures and high fields.
We note that the attractive Hubbard model at $\expv{n}=0.8$, the case
we focused on in this work, corresponds to the $S_z=0.1$ line; it is
nearly perfectly linear for $T>T_K$ and has some weak curvature
much below $T_K$. The impurity is best characterized by its
thermodynamic properties, defined as the impurity contributions to the
total quantities. In panels (b,c,d) we show the results for spin and
charge susceptibility, and the entropy in the $(T,B)$ plane, while
panel (e) presents the spin susceptibility along a set of
constant-magnetization contours. We observe that there are no sharp
features in any of these results: the cross-overs are all smooth, with
no visible kinks. This should be compared with the $\mu$ vs. $T$
curves for the attractive Hubbard model presented in Fig.~\ref{fig1b},
where a kink becomes noticeable for sufficiently negative $U$. Such
kinks must thus be generated through the self-consistency loop and are
a genuine lattice effect that is not present at the single-impurity
level. The susceptibility curves in panel (e) indicate that the
cross-over scale does not depend much on the magnetization. This
property of the pure impurity model explains the results for the
resistivity of the Hubbard model shown in Fig.~\ref{rho} which
indicate an analogous lack of dependence on the band filling.

In Fig.~\ref{siamV2}(a), we show the temperature and field
dependence of the ``conductivity'' for a single spin species of the
symmetric Anderson impurity model as a function of temperature and
magnetic field. The quantity shown is
\begin{equation}
F(T,B) = \int A_\sigma(\omega) \frac{\beta}{4 \cosh(\beta \omega/2)}
\mathrm{d}\omega,
\end{equation}
i.e., the spin-resolved spectral function integrated with a thermal
broadening kernel. A single spin component is considered because under
the partial particle-hole transformation, the original $U<0$ spectral
functions for {\sl both} spins map to a single spin-resolved function
of the $U>0$ model (this is strictly true at the particle-hole
symmetric point). The thermal kernel is the same as in the bulk
expression for the dc conductivity [Eq.~\eqref{sigma} in the
$\Omega\to0$ limit]. If the quantity $F(T,B)$ is evaluated along the
constant-magnetization contours we obtain the results shown in
Fig.~\ref{siamV2}(b): the conductivity is monotonically decreasing,
thus this simple calculation does not explain the nonmonotonous
transport properties of the bulk attractive-$U$ Hubbard model.

One final remark is in order. Fig.~\ref{siam} indicates that there is
nothing special about the zero magnetization line at $B=0$ and that
the results along the zero magnetization contour do not differ
drastically from those for finite magnetization lines. This simply
shows that as the doping in the attractive $U$ Hubbard model is
reduced toward zero, the results are smoothly connected with those for
the repulsive $U$ Hubbard model at half-filling in the absence of the
field, except for the effects of the mapping of spectral functions,
Eq.~\eqref{Amap}, on the transport properties, in particular the
thermopower, as already commented above [Eqs.~\eqref{eq12} and
\eqref{eq13}].

\section{Experimental relevance}
\label{sec8}

\subsection{Zeolites}

\begin{figure*}
\centering
\includegraphics[clip,width=0.98\textwidth]{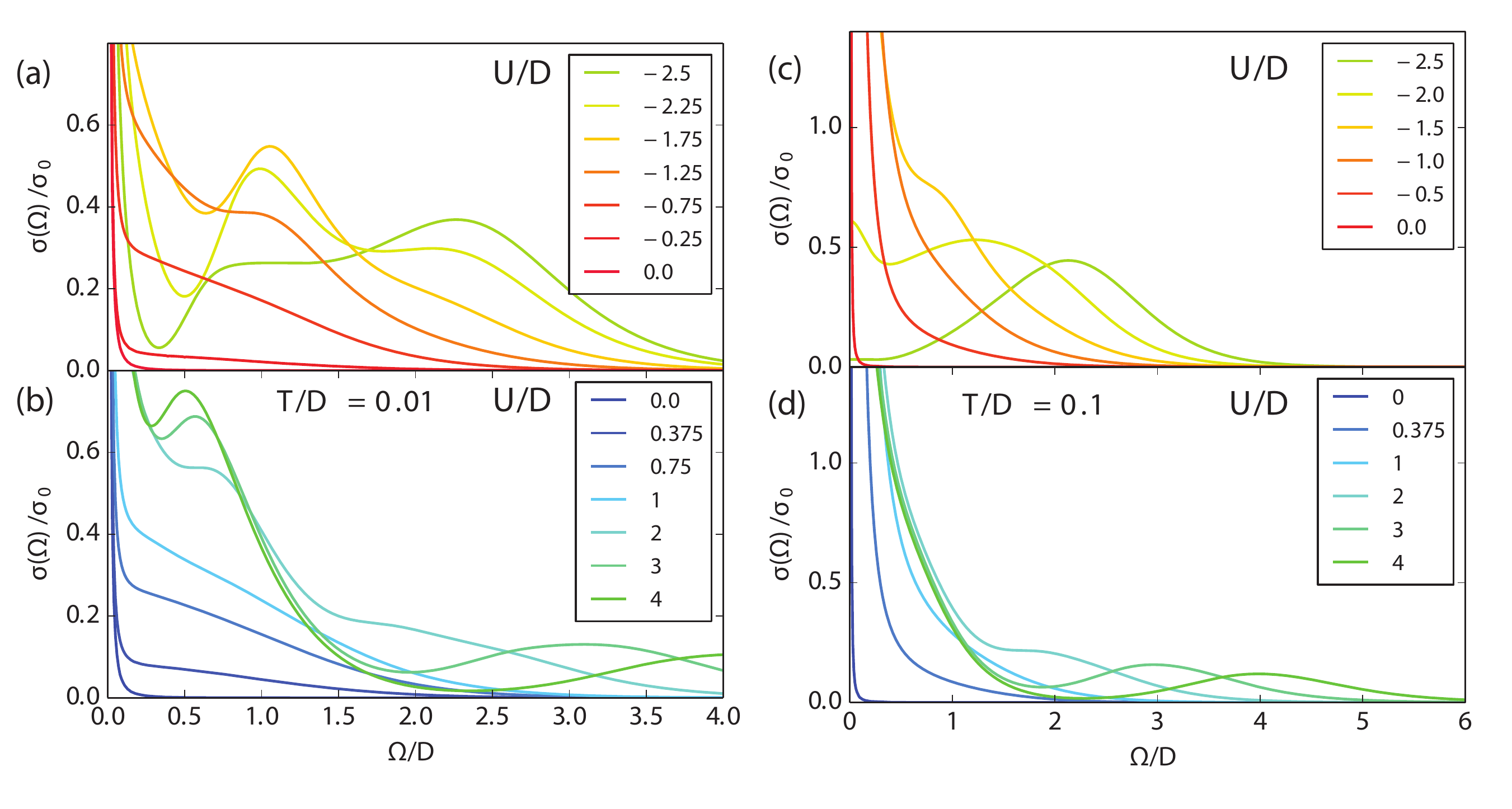}
\caption{(Color online) Optical conductivity for $n=0.8$ for a range
of repulsion parameters $U$. The finite width of the Drude peak for
$U=0$ is due to artificial broadening in the calculation. The absence
of the Drude peak for $U/D=-2.5$ at $T/D=0.1$ shows that the system is
in the (bad) insulator regime.}
\label{figF}
\end{figure*}

Zeolites are aluminosilicate materials with microporous structure
consisting of cages or channels with large voids which can accomodate
alkali cations. They show a variety of exotic electronic properties,
including different magnetically ordered states
\cite{Hanh:2014fw,Nakano:2013es} and metal-insulator transitions
\cite{Nakano:2013es}. The $s$ electrons of alkali atoms are believed
to be confined in the cages and the concentration of dopants strongly
affects the electronic properties, since it changes not only the band
filling, but also the electronic potential depth, thereby controlling
the electron-electron repulsion. Furthermore, the cations can undergo
large displacements, thus there is significant electron-phonon
coupling leading to polaron effects
\cite{Nakano:2013es,PhysRevB.87.075138}. The appropriate model for
such systems is thus some multi-orbital variant of the
Hubbard-Holstein model which takes into account the large number of
electron orbitals inside the cages, and their consecutive filling as
the concentration of dopant atoms is increased. The minimal model,
however, is the single-orbital Hubbard-Holstein model, which may be
expected to describe at least qualitatively the electrons in the
top-most electronic band near the Fermi level. A detailed study of
this model is beyond the scope of the present work. Nevertheless, the
Hubbard-Holstein model maps in the antiadiabatic limit onto the
Hubbard model with effective interaction $U_\mathrm{eff}$ that depends
on the original electron-electron repulsion $U$ and on the value of
the electron-phonon coupling $g$, thus some features of interest can
be studied in this setting. 

A question of direct experimental relevance is how the evolution of
the two key parameters, the band-occupancy $n$ and the coupling $U$,
is reflected in measurable quantities. The optical conductivity for a
range of $n$ at constant $U$ was already shown (Fig.~\ref{figE}) and
here we provide the results for a range of $U$ at constant
$n$ in Fig.~\ref{figF}. The calculations are again performed at
$T/D=0.01$ (left panels) and $T/D=0.1$ (right panels); the lower value
is representative of low-temperature measurements, and the higher one
of those near room temperature. As expected, the variation as a
function of $U$ is much stronger than the dependence on $n$. It
affects the optical conductivity on all frequency scales. At low $U$,
the optical spectrum has a strong Drude peak with a ``Drude
foot''\cite{berthod}, but it is otherwise featureless; a well defined
structure becomes observable only for $|U| \gtrsim D$. Note that in
the true Hubbard-Holstein model we expect a complex optical conductivity
even for $U_\mathrm{eff}=0$, since the effective coupling is itself a
frequency dependent quantity.

\subsection{Optical lattices}

The results of this work are also directly relevant for the
experiments on fermionic cold atoms confined in optical lattices
\cite{Bloch:2008gl}. The value and even the sign of the interparticle
interaction can be tuned at will using the Feshbach resonances
\cite{Chin:2010kl}. Since fermions are difficult to cool down to very
low temperatures (below $0.1E_F$, where $E_F$ is the Fermi energy),
the ordered ground states (quantum magnetism) are not easy to reach
\cite{Greif:2013kb}. For this reason, our results for the paramagnetic
regime above ordering temperatures are actually precisely in the
parameter range accessible to experiments. Recently, experiments
aiming to measure the transport properties have been successfully
performed \cite{Brantut:2012dr,Brantut:2013gs}. Our results on the
Hubbard model will become pertinent once similar experiments are
performed on fermions in optical lattices. Such measurements should be
able to detect the resistivity peak in excess of the MIR limit in the
attractive $U$ case.

\section{Conclusion}

We have compared the basic properties of the Hubbard model constrained
to the paramagnetic phase, with either repulsive or attractive
electron-electron interactions for the same generic value of the
occupancy $\expv{n}=0.8$. The negative-$U$ model can be understood in
terms of the mapping via a partial particle-hole transformation to a
positive-$U$ model at half-filling in external magnetic field such
that the magnetization is fixed to some constant value. This
constraint leads to some interesting features.  The resistivity in the
attractive model strongly increases as the system approaches the
transition to the pairing state (bipolaron formation). There would be
phase separation, signaled in our calculations by the lack of
convergence. The resistivity as a function of the temperature in the
attractive model is non-monotonous: it has a maximum on the scale
$T_\mathrm{max}=ZD$ where the quasiparticles disappear. The NMR
relaxation rate in the attractive model has a complex non-monotonic
temperature dependence which reflects the non-monotonic behavior of
the double occupancy. Since strongly correlated metals with large
electron-phonon coupling can have effective electron-electron
interaction of either sign depending on the system parameters, our
results provide some guidelines to distinguish the repulsive and
attractive interaction in experiments.

\appendix

\section{Spectral sum-rules in the NRG}
\label{appA}

In this appendix we discuss the spectral sum-rules, the
fluctuation-dissipation theorem, and the constraints to their
applicability due to the non-exact nature of the NRG calculations, in
particular at finite $T$, where the density-matrix NRG methods need to
be used \cite{hofstetter2000,anders2005,peters2006,weichselbaum2007}.
The Green's function associated with operators $A$ and $B$ are defined
as \cite{pruschke}
\beq{
G_{AB}(t)=-i\theta(t) \expv{ [A(t),B]_\epsilon },
}
where $\epsilon=+1$ (anti-commutator) if $A$ and $B$ are both
fermionic, and $\epsilon=-1$ (commutator) otherwise. Furthermore, the
correlation functions are defined as
\beq{
\begin{split}
C^>_{AB}(t) &= \expv{A(t)B}, \\
C^<_{AB}(t) &= \expv{BA(t)},
\end{split}
}
and the lesser and greater Green's functions as
\beq{
\begin{split}
G^>_{AB}(t) &= -i\theta(t) \expv{ A(t)B }, \\
G^<_{AB}(t) &= -i\theta(t) \epsilon \expv{ BA(t) },
\end{split}
}
The Fourier transforms are 
\beq{
C^{>,<}(\omega) = \int_{-\infty}^{\infty} \mathrm{d}t\, e^{i\omega t}
C^{>,<}(t),
}
and the Laplace transform of the Green's functions ($z=\omega+i0^+$) is
\beq{
G_{AB}(z) = \int_0^\infty \mathrm{d}t\, e^{izt} G_{AB}(t).
}

The relation between $C^{>,<}$ and $G^{<,>}$ is
\beq{
\begin{split}
[G^>_{AB}]''(\omega) = -\pi C^>_{AB}(\omega),
\\
[G^<_{AB}]''(\omega) = -\pi \epsilon C^<_{AB}(\omega).
\end{split}
}
Here $G''(\omega)$ denotes the jump funciton, which is here
equal to the imaginary part of retarded Green's function, i.e.,
$\Im\,G(\omega+i\delta)$.
The total spectral function can be written in several equivalent
forms:
\beq{
\begin{split}
\rho_{AB}(\omega) &= C^>_{AB}(\omega) + \epsilon C^<_{AB}(\omega) \\
&= -\frac{1}{\pi} \left\{ [G_{AB}^>]''(\omega+i0) +
[G_{AB}^<]''(\omega+i0) \right\} \\
&= -\frac{1}{2\pi i} \left[ G_{AB}(\omega+i0) - G_{AB}(\omega-i0)
\right] \\
&= -\frac{1}{\pi} G''_{AB}(\omega+i0).
\end{split}
}

Using Lehmann's decomposition, one can show that
\beq{
\label{A8}
C^>_{AB}(\omega) e^{-\beta \omega} = C^<_{AB}(\omega),
}
thus 
\beq{
C^>_{AB}(\omega) = \frac{G''_{AB}(\omega)}{1+\epsilon e^{-\beta \omega}}.
}
From this one obtains 
\beq{
\expv{A(t)B} = \int_{-\infty}^{\infty} \mathrm{d}\omega\, e^{-i\omega
t} \frac{1}{\pi} \frac{G''_{AB}(\omega+i0)}
{1+\epsilon e^{-\beta \omega}},
}
and finally the fluctuation-dissipation theorem (FDT) in the form
\beq{
\label{fdt}
\expv{AB} = \int_{-\infty}^{\infty} \mathrm{d}\omega \frac{\rho_{AB}(\omega)}
{1+\epsilon e^{-\beta \omega}}.
}
Alternatively, by integrating over the $C^{<,>}$ functions, one
can obtain
\beq{
\begin{split}
\int_{-\infty}^{\infty} C^>_{AB}(\omega)\mathrm{d}\omega &=
\expv{AB}, \\
\int_{-\infty}^{\infty} C^<_{AB}(\omega)\mathrm{d}\omega &= \expv{BA}.
\end{split}
}
It turns out that in the full-density-matrix numerical renormalization
group (FDM-NRG), these two sum-rules are satisfied exactly by
construction (up to floating-point round-off errors of order
$10^{-16}$), as long as the expectation values on the right-hand-side
are evaluated using the suitable density-matrix aproach
\cite{weichselbaum2007}. This is not the case, however, for the FDT in
the form of Eq.~\eqref{fdt}. It turns out that there is nothing in the
NRG that guarantees that the detailed balance relation
$C^>_{AB}(\omega) e^{-\beta \omega} = C^<_{AB}(\omega)$,
Eq.~\eqref{A8}, should be fullfilled by construction. Greater and
lesser correlation functions are calculated somewhat differently
because in the FDM-NRG the expansions of the identity into kept and
discarded states need to be performed differently in each case. In
practice, at $T=0$ the FDT from Eq.~\eqref{fdt} is fulfilled to
numerical precision, but the error grows with increasing $T$. At very
high temperature $T=0.1D$, for example, the violation of the FDT is
about one permil for the fermionic spectral function and a few percent
for the dynamical spin susceptibility. This implies that the sum-rules
need to be checked at the level of $C^>$ and $C^<$ correlation
functions.

\begin{acknowledgments}
We acknowledge very useful discussions with J. Mravlje. R. \v{Z}. and
\v{Z}. O. acknowledge the support of the Slovenian Research Agency
(ARRS) under Program No. P1-0044.
\end{acknowledgments}

\bibliography{paper}

\end{document}